\documentclass[11pt, a4paper]{article}
\usepackage{graphicx}
\usepackage{placeins}
\usepackage{float}
\usepackage{caption}
\usepackage{setspace}
\usepackage{hanging}
\usepackage[utf8]{inputenc}
\usepackage[left=3cm,right=3cm, bottom=2.5cm, top=3cm]{geometry}
\usepackage{amssymb, amsmath, amssymb, amsthm}
\usepackage{color}
\usepackage{natbib}
\usepackage{graphicx}
\usepackage{xcolor}
\usepackage{tabularx}
\usepackage{threeparttable}
\usepackage{longtable}
\usepackage{mathdots}
\usepackage{setspace}
\usepackage{booktabs}
\usepackage{caption}
\usepackage{subcaption}
\usepackage{makecell}
\usepackage{longtable}
\usepackage{makecell}
\usepackage{tikz}
\usepackage{rotating}
 \usepackage[nice]{nicefrac}
\usepackage{soul} 
\usepackage{pgf}

\usepackage	[colorlinks,
    linkcolor=blue,
	citecolor=blue,
	urlcolor=blue]{hyperref}
\usepackage{siunitx}
\usepackage[title]{appendix}

\usepackage[sc]{mathpazo} 
\usepackage{newpxmath}

\usepackage{tikz}
\usetikzlibrary{shapes,decorations,arrows,calc,arrows.meta,fit,positioning}
\tikzset{
	-Latex,auto,node distance =1 cm and 1 cm,semithick,
	state/.style ={ellipse, draw, minimum width = 0.7 cm},
	point/.style = {circle, draw, inner sep=0.04cm,fill,node contents={}},
	bidirected/.style={Latex-Latex,dashed},
	el/.style = {inner sep=2pt, align=left, sloped}
}

\pgfmathsetmacro{\AggregateUpliftingLB}{58}
\pgfmathsetmacro{\AggregateUpliftingUB}{97}
\pgfmathsetmacro{\ForestsUpliftingLB}{20}
\pgfmathsetmacro{\ForestsUpliftingMed}{40}
\pgfmathsetmacro{\ForestsUpliftingUB}{65}
\pgfmathsetmacro{\UpliftingForestsAtThirty}{47}
\pgfmathsetmacro{\UpliftingForestsAtOneHundred}{260}
\pgfmathsetmacro{\UpliftingDifference}{100*(\AggregateUpliftingLB-\ForestsUpliftingMed)/(\ForestsUpliftingMed)}

\pgfmathsetmacro{\CountWTPStudies}{396}
\pgfmathsetmacro{\CountWTPEstimates}{735}
\pgfmathsetmacro{\MainEstimate}{0.587111} 
\pgfmathsetmacro{\MainEstimateSE}{0.0694836}
\pgfmathsetmacro{\UnivariateEstimate}{0.3792949}
\pgfmathsetmacro{\UnivariateEstimateSE}{0.066298}
\pgfmathsetmacro{\AverageBasisEstimate}{0.58}
\pgfmathsetmacro{\AllSeparateBasisEstimate}{0.64}
\pgfmathsetmacro{\CountAllSeparateBasisEstimate}{851}
\pgfmathsetmacro{\OnlyPositiveWTPEstimates}{0.61}
\pgfmathsetmacro{\ClimateregulationEst}{0.7572566}
\pgfmathsetmacro{\ClimateregulationSE}{0.0941662}
\pgfmathsetmacro{\ClimateregulationN}{165}
\pgfmathsetmacro{\ClimateregulationRsquared}{0.63}
\pgfmathsetmacro{\AirqualityregulationEst}{0.5221629}
\pgfmathsetmacro{\AirqualityregulationSE}{0.1710705}
\pgfmathsetmacro{\AirqualityregulationN}{186}
\pgfmathsetmacro{\AirqualityregulationRsquared}{0.44}
\pgfmathsetmacro{\WaterregulationEst}{0.5108777}
\pgfmathsetmacro{\WaterregulationSE}{0.1308559}
\pgfmathsetmacro{\WaterregulationN}{212}
\pgfmathsetmacro{\WaterregulationRsquared}{0.59}
\pgfmathsetmacro{\NaturalhazardregulationEst}{0.5174335}
\pgfmathsetmacro{\NaturalhazardregulationSE}{0.1910137}
\pgfmathsetmacro{\NaturalhazardregulationN}{72}
\pgfmathsetmacro{\NaturalhazardregulationRsquared}{0.84}
\pgfmathsetmacro{\ErosionregulationEst}{0.8441836}
\pgfmathsetmacro{\ErosionregulationSE}{0.1160251}
\pgfmathsetmacro{\ErosionregulationN}{125}
\pgfmathsetmacro{\ErosionregulationRsquared}{0.74}
\pgfmathsetmacro{\WaterpurificationregulationEst}{0.4342403}
\pgfmathsetmacro{\WaterpurificationregulationSE}{0.201589}
\pgfmathsetmacro{\WaterpurificationregulationN}{95}
\pgfmathsetmacro{\WaterpurificationregulationRsquared}{0.48}
\pgfmathsetmacro{\RegulatingServicesEst}{0.6045562}
\pgfmathsetmacro{\RegulatingServicesSE}{0.0730538}
\pgfmathsetmacro{\RegulatingServicesN}{445}
\pgfmathsetmacro{\RegulatingServicesRsquared}{0.40}
\pgfmathsetmacro{\SpiritualandreligiousvaluesEst}{0.8703952}
\pgfmathsetmacro{\SpiritualandreligiousvaluesSE}{0.186381}
\pgfmathsetmacro{\SpiritualandreligiousvaluesN}{57}
\pgfmathsetmacro{\SpiritualandreligiousvaluesRsquared}{"-"}
\pgfmathsetmacro{\AestheticvaluesEst}{0.626019}
\pgfmathsetmacro{\AestheticvaluesSE}{0.080599}
\pgfmathsetmacro{\AestheticvaluesN}{341}
\pgfmathsetmacro{\AestheticvaluesRsquared}{0.47}
\pgfmathsetmacro{\RecreationandecotourismEst}{0.7364278}
\pgfmathsetmacro{\RecreationandecotourismSE}{0.0898021}
\pgfmathsetmacro{\RecreationandecotourismN}{338}
\pgfmathsetmacro{\RecreationandecotourismRsquared}{0.51}
\pgfmathsetmacro{\BiodiversityEst}{0.7697451}
\pgfmathsetmacro{\BiodiversitySE}{0.0947553}
\pgfmathsetmacro{\BiodiversityN}{343}
\pgfmathsetmacro{\BiodiversityRsquared}{0.61}
\pgfmathsetmacro{\CulturalServicesEst}{0.6472808}
\pgfmathsetmacro{\CulturalServicesSE}{0.0863148}
\pgfmathsetmacro{\CulturalServicesN}{433}
\pgfmathsetmacro{\CulturalServicesRsquared}{0.45}
\pgfmathsetmacro{\ForestEst}{0.6581393}
\pgfmathsetmacro{\ForestSE}{0.141571}
\pgfmathsetmacro{\ForestN}{177}
\pgfmathsetmacro{\ForestRsquared}{0.69}
\pgfmathsetmacro{\NonForestEst}{0.6250036}
\pgfmathsetmacro{\NonForestSE}{0.0818389}
\pgfmathsetmacro{\NonForestN}{558}
\pgfmathsetmacro{\NonForestRsquared}{0.43}
\pgfmathsetmacro{\NorthAmericaEst}{0.86} 
\pgfmathsetmacro{\NorthAmericaSE}{0.72}
\pgfmathsetmacro{\NorthAmericaN}{88}
\pgfmathsetmacro{\NorthAmericaRsquared}{0.71}
\pgfmathsetmacro{\SouthAmericaEst}{-0.07} 
\pgfmathsetmacro{\SouthAmericaSE}{0.62}
\pgfmathsetmacro{\SouthAmericaN}{37}
\pgfmathsetmacro{\SouthAmericaRsquared}{"-"}
\pgfmathsetmacro{\AfricaEst}{0.80}
\pgfmathsetmacro{\AfricaSE}{0.17}
\pgfmathsetmacro{\AfricaN}{38}
\pgfmathsetmacro{\AfricaRsquared}{"-"}
\pgfmathsetmacro{\EuropeEst}{0.82}
\pgfmathsetmacro{\EuropeSE}{0.13}
\pgfmathsetmacro{\EuropeN}{269}
\pgfmathsetmacro{\EuropeRsquared}{0.56}
\pgfmathsetmacro{\AsiaEst}{0.28} 
\pgfmathsetmacro{\AsiaSE}{0.11}
\pgfmathsetmacro{\AsiaN}{300}
\pgfmathsetmacro{\AsiaRsquared}{0.51}
\pgfmathsetmacro{\FirstPeriodEst}{0.65}
\pgfmathsetmacro{\FirstPeriodSE}{0.11}
\pgfmathsetmacro{\FirstPeriodN}{400}
\pgfmathsetmacro{\FirstPeriodRsquared}{0.29}
\pgfmathsetmacro{\SecondPeriodEst}{0.53}
\pgfmathsetmacro{\SecondPeriodSE}{0.10}
\pgfmathsetmacro{\SecondPeriodN}{301}
\pgfmathsetmacro{\SecondPeriodRsquared}{0.58}
\pgfmathsetmacro{\BelowMedianEst}{0.68}
\pgfmathsetmacro{\BelowMedianSE}{0.07}
\pgfmathsetmacro{\BelowMedianN}{367}
\pgfmathsetmacro{\BelowMedianRsquared}{0.35}
\pgfmathsetmacro{\AboveMedianEst}{0.71}
\pgfmathsetmacro{\AboveMedianSE}{0.21}
\pgfmathsetmacro{\AboveMedianN}{368}
\pgfmathsetmacro{\AboveMedianRsquared}{0.52}
\pgfmathsetmacro{\BottomQuarterEst}{0.67}
\pgfmathsetmacro{\BottomQuarterSE}{0.12}
\pgfmathsetmacro{\BottomQuarterN}{183}
\pgfmathsetmacro{\BottomQuarterRsquared}{0.34}
\pgfmathsetmacro{\TopQuarterEst}{1.91}
\pgfmathsetmacro{\TopQuarterSE}{0.54}
\pgfmathsetmacro{\TopQuarterN}{183}
\pgfmathsetmacro{\TopQuarterRsquared}{0.67}
\pgfmathsetmacro{\SpecChartMainEstimatePercentile}{"63rd"}
\pgfmathsetmacro{\SpecChartMinimumEstimate}{0.35}
\pgfmathsetmacro{\SpecChartMaximumEstimate}{0.67}
\pgfmathsetmacro{\GrowthDiffGDPcapitaRegSvcsEst}{2.83}
\pgfmathsetmacro{\GrowthDiffGDPcapitaRegSvcsSE}{0.17}
\pgfmathsetmacro{\GrowthDiffGDPcapitaCulSvcsEst}{2.95}
\pgfmathsetmacro{\GrowthDiffGDPcapitaCulSvcsSE}{0.09}
\pgfmathsetmacro{\GrowthDiffGDPcapitaAggEst}{2.83}
\pgfmathsetmacro{\GrowthDiffGDPcapitaAggSE}{0.17}
\pgfmathsetmacro{\GrowthDiffGDPcapitaForestsEst}{1.94}
\pgfmathsetmacro{\GrowthDiffGDPcapitaForestsSE}{0.07}
\pgfmathsetmacro{\RegulatingServicesME}{ ( (\RegulatingServicesSE/(\RegulatingServicesEst))^2 + (\GrowthDiffGDPcapitaRegSvcsSE/(\GrowthDiffGDPcapitaRegSvcsEst))^2 )^(1/2) }
\pgfmathsetmacro{\RPCRegulatingServices}{\RegulatingServicesEst*\GrowthDiffGDPcapitaRegSvcsEst}
\pgfmathsetmacro{\RPCRegulatingServicesLB}{ \RPCRegulatingServices - (1.96*\RegulatingServicesME) }
\pgfmathsetmacro{\RPCRegulatingServicesUB}{ \RPCRegulatingServices + (1.96*\RegulatingServicesME) }
\pgfmathsetmacro{\CulturalServicesME}{ ( (\CulturalServicesSE/(\CulturalServicesEst))^2 + (\GrowthDiffGDPcapitaCulSvcsSE/(\GrowthDiffGDPcapitaCulSvcsEst))^2 )^(1/2) }
\pgfmathsetmacro{\RPCCulturalServices}{\CulturalServicesEst*\GrowthDiffGDPcapitaCulSvcsEst}
\pgfmathsetmacro{\RPCCulturalServicesLB}{ \RPCCulturalServices - (1.96*\CulturalServicesME) }
\pgfmathsetmacro{\RPCCulturalServicesUB}{ \RPCCulturalServices + (1.96*\CulturalServicesME) }
\pgfmathsetmacro{\AggregateServicesME}{ ( (\MainEstimateSE/(\MainEstimate))^2 + (\GrowthDiffGDPcapitaAggSE/(\GrowthDiffGDPcapitaAggEst))^2 )^(1/2) }
\pgfmathsetmacro{\RPCAggregateServices}{\MainEstimate*\GrowthDiffGDPcapitaAggEst}
\pgfmathsetmacro{\RPCAggregateServicesLB}{ \RPCAggregateServices - (1.96*\AggregateServicesME) }
\pgfmathsetmacro{\RPCAggregateServicesUB}{ \RPCAggregateServices + (1.96*\AggregateServicesME) }
\pgfmathsetmacro{\ForestServicesME}{ ( (\ForestSE/(\ForestEst))^2 + (\GrowthDiffGDPcapitaForestsSE/(\GrowthDiffGDPcapitaForestsEst))^2 )^(1/2) }
\pgfmathsetmacro{\RPCForestServices}{\ForestEst*\GrowthDiffGDPcapitaForestsEst}
\pgfmathsetmacro{\RPCForestServicesLB}{ \RPCForestServices - (1.96*\ForestServicesME) }
\pgfmathsetmacro{\RPCForestServicesUB}{ \RPCForestServices + (1.96*\ForestServicesME) }
\pgfmathsetmacro{\MainEstimateLB}{\MainEstimate - 1.96 * \MainEstimateSE}
\pgfmathsetmacro{\MainEstimateUB}{\MainEstimate + 1.96 * \MainEstimateSE}
\pgfmathsetmacro{\ElasticityOfSubstitutability}{1/\MainEstimate}
\pgfmathsetmacro{\ElasticityOfSubstitutabilityLB}{1/\MainEstimateUB}
\pgfmathsetmacro{\ElasticityOfSubstitutabilityUB}{1/\MainEstimateLB}
\pgfmathsetmacro{\UnivariateEstimateLB}{\UnivariateEstimate - 1.96 * \UnivariateEstimateSE}
\pgfmathsetmacro{\UnivariateEstimateUB}{\UnivariateEstimate + 1.96 * \UnivariateEstimateSE}
\title{Global evidence on the income elasticity of\\[0.2cm] willingness to pay, relative price changes\\[0.2cm] and public natural capital values}
\vspace{-0.5cm}
\author{} 
\author{Moritz A. Drupp$^{\text{a,b}}$,\, Zachary M. Turk$^{\text{c}}$,\\[0.1cm] Ben Groom$^{\text{d,e}}$,\, Jonas Heckenhahn$^{\text{f}}$\footnote{We thank Jasper Meya, Sjak  Smulders, Daan van Soest and Martin Quaas as well as seminar audiences at BIOECON 2023, the World Bank, idiv Leipzig, MWLR and EAERE 2023 for helpful discussions, and are grateful to Johanna Darmstadt, Mark Lustig and Jasper R\"oder for excellent research assistance. We gratefully acknowledge funding by the World Bank. M.D.\ acknowledges support from the German Federal Ministry of Education and Research (BMBF) under grant number 01UT2103B. B.G. acknowledges Dragon Capital for funding the Dragon Capital Chair and funding from the UKRI/NERC BIOADD project (ref: NE/X002292/1). J.H.  acknowledges support from the Evangelisches Studienwerk e.V.\ Villigst. All authors declare that they have no relevant or material financial interests related to the research in this paper. }\\[0.4cm]
\small
$^{\text{a}}$ Department of Economics, University of Hamburg, Germany\\
\small
$^{\text{b}}$ Department of Economics, University of Gothenburg, Sweden\\
\small
$^{\text{c}}$ Department of Economics, Cork University Business School, University College Cork, Ireland\\
\small
$^{\text{d}}$ Department of Economics, University of Exeter Business School, United Kingdom\\
\small
$^{\text{e}}$ Grantham Research Institute on Climate Change and the Environment,\\
\small
London School of Economics and Political Science, UK\\
\small
\small
$^{\text{f}}$ Faculty of Management and Economics, Ruhr-University Bochum, Germany\\
}

\date{\today}

\begin{document}

\maketitle
\thispagestyle{empty}
\onehalfspacing
  \vspace{-1cm}
\begin{abstract}

While the global economy continues to grow, ecosystem services tend to stagnate or decline. Economic theory has shown how such shifts in relative scarcities can be reflected in project appraisal and accounting, but empirical evidence has been sparse to put theory into practice. To estimate relative price changes in ecosystem services to be used for making such adjustments, we perform a global meta-analysis of contingent valuation studies to derive income elasticities of marginal willingness to pay (WTP) for ecosystem services to proxy the degree of limited substitutability. Based on {\CountWTPEstimates} income-WTP pairs from {\CountWTPStudies} studies, we find an income elasticity of WTP of around \num[round-mode=places,round-precision=1]{\MainEstimate}. Combined with good-specific growth rates, we estimate relative price change of ecosystem services of around \num[round-mode=places,round-precision=1]{\RPCAggregateServices} percent per year. In an application to natural capital valuation of forest ecosystem services by the World Bank, we show that natural capital should be uplifted by around {\ForestsUpliftingMed} percent. Our assessment of aggregate public natural capital yields a larger value adjustment of between {\AggregateUpliftingLB} and {\AggregateUpliftingUB} percent, depending on the discount rate. We discuss implications for policy appraisal and for estimates of natural capital in comprehensive wealth accounts.

\end{abstract}
 
\noindent \textbf{Keywords:} Willingness to Pay;  Ecosystem Services; Income Elasticity; Limited Substitutability; Growth;  Relative Prices; Contingent Valuation; Forests; Natural Capital
 
\smallskip
\noindent \textbf{JEL codes:} D61, H43, Q51, Q54, Q58

\newpage
\section{Introduction}

Measuring economic progress towards sustainability requires addressing the limited substitutability among the various constituents of comprehensive wealth \citep{smulders2023}. Potential limits to substitution imply that society must strike a balance between the two opposing paradigms of Weak and Strong Sustainability \citep[e.g.,][]{neumayer2003weak, hanleyetal2015, dasgupta2021economics}. Many contemporary measures of economic progress and wealth have explicitly or implicitly followed a Weak Sustainability approach. In doing so, they consider natural capital and ecosystem services as largely substitutable---sometimes even perfectly substitutable---with human-made capital stocks. 
In light of the continued growth of human-made capital and the stagnation or degradation of many natural capital stocks \citep {IPBES2019}, the Weak Sustainability approach is increasingly being called into question. From a theory perspective, we should consider some degree of imperfect substitutability when estimating shadow prices. This is relevant both for natural capital that serves as an intermediate input to various production processes and for public natural capital as a direct source of utility \citep[see, e.g.][]{smulders2023, zhu2019discounting}. A common constraint to implementation, however, has been a lack of sufficient empirical evidence on the degree of substitutability of ecosystem services and natural capital to inform the computation of shadow prices \citep[e.g.,][]{cohen2019natural, drupp2018limits, drupp2024accounting, rouhi2021complementarity}. 

This paper makes a step towards closing this important empirical evidence gap by characterising the limited degree of substitutability of ecosystem services in utility via a global meta-analysis of contingent valuations (CV) studies. Doing so allows changes in the relative scarcity of ecosystem services to be more appropriately valued in policy appraisal and environmental-economic accounting. The evidence is drawn from the largest global meta-analysis to date that estimates  income elasticities of marginal willingness to pay (WTP) for ecosystem services as a proxy for the limited substitutability of ecosystem services vis-a-vis market goods. Knowing the income elasticity of WTP, and good-specific growth rates that we estimate as well, allows computing the relative price changes of ecosystem services. We then propose an approach to deriving adjustments to natural capital accounts, which we calibrate with our empirical estimates for the case of non-wood forest ecosystem service values in the  World Bank's \textit{Changing Wealth of Nations (CWON)} measure and for an aggregate  adjustment of public natural capital values.

There are two general approaches to reflecting limited substitutability of ecosystem services in Cost-Benefit Analysis (CBA) or the assessment of comprehensive wealth, to take two examples. One can either apply differentiated discount rates---often a lower discount rate for non-market ecosystem services---or account for increasing relative scarcity by adjusting our valuation (accounting price) of ecosystem services throughout the horizon of the evaluation \citep[e.g.][]{baumgartner2015ramsey, drupp2018limits, gollier2010ecological, hoel2007discounting,  traeger2011sustainability, weikard2005discounting}. Several studies have already shown the importance of accounting for the adverse effects of climate change on ecosystem services, biodiversity and environmental amenities \citep[e.g.][]{hoel2007discounting, sterner2008even}. More recently \cite{drupp2021relative} and \cite{bastien2021use} examined how the increasing scarcity and limited substitutability of non-market ecosystem services each affect optimal climate policy through good-specific discount rates or relative price changes. \cite{drupp2021relative}, for instance, estimate that limited substitutability leads to relative prices of non-market goods increasing by around 2 to 4 percent per year. Incorporating this scale of adjustment leads to estimates of the social cost of carbon that are more than 50 percent higher compared to the case where goods are assumed to be perfectly substitutable. Accounting for relative price changes of non-market goods is thus crucial to the appraisal of climate policy. Perhaps more importantly, relative price changes need to be accounted for properly in the appraisal of projects, regulations, and policies to better account for the impact of ecosystem services on well-being. Furthermore, when using environmental-economic accounting, e.g. within the UN System of Environmental Economic Accounting-Experimental Ecosystem Accounting (SEEA-EEA) or \textit{CWON}, valuations that account for limited substitutability are critical to the assessment of sustainability.

Practically speaking, two components are needed to estimate the trajectory of relative prices for ecosystem services: First, the elasticity of substitution between market and non-market goods; and, second, their respective growth rates. Previous empirical studies have estimated the elasticity of substitution indirectly using the inverse of the income elasticity of willingness to pay (WTP) from non-market valuation studies \citep{baumgartner2015ramsey, drupp2018limits, heckenhahn2024relative}. Good-specific growth rates have been estimated either using historical time series data \citep[e.g.,][]{baumgartner2015ramsey, heckenhahn2024relative}, and then assuming constant exponential growth rates as we will do our main application here, or as endogenous outcomes in global integrated climate-economy assessment models \citep[e.g.][]{sterner2008even, drupp2021relative, bastien2021use}. The rate of change of relative prices is then approximated by the income elasticity of WTP multiplied by the difference between the growth rates of marketed and non-marketed goods. \cite{baumgartner2015ramsey} were the first to estimate relative price changes in this way.  Yet, the study drew on an estimate of the elasticity of substitution for just one ecosystem service: global biodiversity conservation, based on a small meta-analysis of 46 CV studies by \cite{jacobsen2009there}. \cite{heckenhahn2024relative} provide the first country-specific evidence for Germany, which built on just 36 WTP studies. There is thus clear room for improvement in the estimates of growth and substitutability of ecosystem services so that welfare effects and sustainability can be more accurately evaluated.

These gaps in and limitations of the empirical evidence---the absence of both a general default for generic ecosystem services as well as ecosystem service-specific estimates of income or substitution elasticities and growth rates---mean that guidelines for governmental appraisal and environmental-economic accounting only rarely address the issue of  limited substitutability of non-market goods \citep{groom2022future}. 
Where environmental discounting or relative price changes have been integrated into governmental policy guidance, they are operationalized using very coarse estimates of growth rates and elasticities  \citep{groom2017reflections}. For instance, The Netherlands consider a general default relative price change of 1 percent per annum for ecosystem services of all kinds in their discounting guidance, following the estimate based solely on biodiversity related services from \cite{baumgartner2015ramsey}.\footnote{The guidance allows specific deviations from 1\% if growth or substitution possibilities deviate from the default assumptions, e.g. if the ecosystem service is deemed no-substitutable.} 
The UK Department for Environment, Food and Rural Affairs used to reflect relative price adjustments for the health benefits of pollution reductions by `uplifting’ the damage costs by 2 percent per year. The underlying assumption here is that WTP for avoiding the health consequences of pollution grows in line with predicted income \citep{HMT2021_Environmental}. Indeed, health benefits in general are discounted using a discount rate that is 2 percentage points lower in the UK for related reasons.\footnote{Another rationale for this practice stems from the use of Quality Adjusted Life Years in UK public appraisal. Since QALYs are measured in utility, so the argument goes, they should only be discounted using the pure rate of time preference} For the environment in general, where the guidelines do reflect changing valuations over time or lower discount rates, e.g. in the Asian Development Bank and Canadian guidelines, again rather generic rules of thumb are used that do not distinguish across different ecosystem services \citep{groom2022future}. Finally, where guidelines exist for natural capital valuation, such as the 2021  \textit{CWON} report by the World Bank, they apply to a minimal basket of non-market goods and capital stocks, and fail to account for changing relative prices. For instance, in the CWON forest ecosystem services are valued using a meta-regression and spatial benefit transfer by \cite{siikamaki2015global}, yet maintain constant real prices over time and thereby fail to complement spatial benefit transfer with intertemporal benefit transfer.\footnote{The assumption is that per-hectare monetary values are constant over time (correcting for inflation). Note, \cite{siikamaki2015global} find positive and large GDP elasticities of WTP.} 

From a policy perspective, not accounting for limited substitutability of ecosystem services and market goods, either via relative price changes or in discount rates, means that ecosystem services will be seriously undervalued in public appraisal of policy or natural capital, particularly if relative scarcity is rising. The underlying---often implicit---assumption in such cases is that ecosystem services are perfectly substitutable with market goods. Yet, even in the unusual cases where adjustments have been made, the advice is too generic to properly reflect sustainability and the welfare associated with different ecosystem services over time. For practical purposes, then, more accurate estimates are needed, ideally differentiated across ecosystem services in case sizable heterogeneities manifest themselves. 

Against this background, we provide first systematic, global empirical evidence basis to inform relative price adjustments of ecosystem services---both for a proxy of aggregate ecosystem services as well as for a ecosystem services sub-categories. These estimates can be applied to public appraisal of public investment and regulatory change as well as to natural capital valuation such as the \textit{CWON} program and the SEEA-EEA.  
Our main focus is on improving the estimation of limited substitutability of non-market ecosystem services vis-a-vis market goods. To this end, we perform a meta-analysis of environmental values derived using the CV method to estimating the income elasticity of WTP---a key parameter also for benefit transfer across space \citep{baumgartner2017income, smith2023accounting}.  Our meta-analysis draws on a large-scale keyword-based search strategy and an in-depth analysis of the known population of more than 2000 peer-reviewed CV studies. 
Our full sample includes {\CountWTPEstimates} mean income and WTP estimates, including recurring covariates, sourced from {\CountWTPStudies} peer-reviewed CV  studies.

Our central estimate suggest an income elasticity of WTP for ecosystem services on aggregate of about \num[round-mode=places,round-precision=1]{\MainEstimate}, with a 95 confidence interval extending from around \num[round-mode=places,round-precision=1]{\MainEstimateLB} to \num[round-mode=places,round-precision=1]{\MainEstimateUB}. Point estimates for different ecosystem service types range between about \num[round-mode=places,round-precision=1]{\WaterpurificationregulationEst} (water purification and waste treatment) and \num[round-mode=places,round-precision=1]{\SpiritualandreligiousvaluesEst} (spiritual and religious values). Using estimates of good-specific growth rates, we compute relative price changes of ecosystem services of around \num[round-mode=places,round-precision=1]{\RPCAggregateServices} percent per year on aggregate. Relative price changes are smaller for forest ecosystem services (\num[round-mode=places,round-precision=1]{\RPCForestServices} percent), primarily due to a lower rate of de-growth of forest area. These estimates can be employed to adjust WTP estimates for project appraisal or environmental-economic accounting. In an application on natural capital valuation, taking the \textit{CWON} 2021 report by the \cite{world2021changing}, we show that adjusting natural capital estimates for non-timber ecosystem services for relative price changes results in uplifting the present value over a 100-year time period by {\ForestsUpliftingMed} percent (95 CI: {\ForestsUpliftingLB} to {\ForestsUpliftingUB} percent), materially elevating the role of public natural capital. Our estimates for adjustments to the value of aggregate public natural capital are more substantial, amounting to between {\AggregateUpliftingLB} and {\AggregateUpliftingUB} percent for our main estimate, depending on the social discount rate. These results echo work on the importance of limited substitutability in climate policy appraisal \citep{bastien2021use, drupp2021relative, sterner2008even}. We close by discussing limitations of our analysis and by summarizing insights for project appraisal, accounting, and sustainability more generally.

\bigskip
\section{Theoretical background}

To provide the theoretical background for our empirical analysis, we consider a simple model in which intertemporal well-being is derived from both human-made goods, $C_t$ and non-market environmental goods or ecosystem services, $E_t$.  In the general case of imperfect substitutability, ecosystem services feature explicitly in the instantaneous utility function representing preferences over market-traded consumption goods and non-market goods, $U({C_t,E_t})$. A standard form of the time-discounted Utilitarian social welfare function is given by:
\begin{equation}
W=\int_{t=0}^{\infty}U({C_t,E_t})e^{-\delta t} dt \, . 
\end{equation}

The theory of dual discounting or relative price changes has shown that there are two approaches to addressing the intertemporal appraisal of non-market goods \citep[e.g., ][]{baumgartner2015ramsey, gollier2010ecological, traeger2011sustainability, weikard2005discounting}: 
\begin{enumerate}
\item Explicitly consider how the relative price of non-market goods vis-a-vis market-traded consumption goods changes over time. Then, compute comprehensive consumption equivalents at each point in time and use a single consumption discount rate to on future comprehensive consumption equivalents.
\item Use differentiated, good-specific consumption discount rates, i.e.\ one for market goods, $r_C$, and another for non-market goods, $r_E$. 
\end{enumerate}

In the first approach, we compute the value of non-market goods in terms of the market good numeraire. This value is given by the marginal rate of substitution (MRS), $U_{E_t} / U_{C_t}$, which is the implicit price of non-market goods. The MRS tells us by how much the consumption of market goods would need to increase in response to a marginal decrease in non-market goods to hold utility constant. The $RPC_t$ measures the change in the MRS between non-market and market goods over time, i.e.\ the relative change in the valuation of non-market goods \citep{hoel2007discounting}:
\begin{equation}
\label{Equ_RelPriceGeneral}
RPC_t = 	\nicefrac{\dfrac{d}{dt} \left( \dfrac{U_{E_t}}{U_{C_t}}\right)}{\left( \dfrac{U_{E_t}}{U_{C_t}}\right)} \, .
\end{equation}  

Future expected non-market values can then be adjusted using the $RPC_t$ and a single SDR can then be used to discount future flows of private and non-market consumption.

In the second approach, we compute good-specific (dual) discount rates as:
\begin{equation}
r_{C_t} = \delta + \eta_{CC_t} g_{C_t} + \eta_{CE_t} g_{E_t}
\end{equation}
\begin{equation}
r_{E_t} = \delta + \eta_{EE_t} g_{E_t} + \eta_{EC_t} g_{C_t} 
\end{equation}
where $g_E$ and $g_C$  are the growth rates, $\eta_{CC_t}$ ($\eta_{EE_t}$) is the elasticity of marginal utility of private-good (non-market good) consumption with respect to private-good (non-market good) consumption, and $\eta_{CE_t}$ ($\eta_{EC_t}$) denotes the cross-elasticity of marginal utility of private-good (non-market good) consumption with respect to non-market good (private-good) consumption \citep[see, e.g., ][]{baumgartner2015ramsey}. Expanding their applicability, these dual rates can also be used in cases where non-market goods are not evaluated in monetary units such as satellite accounts in national accounting and biophysical impact assessments. It is important to stress that this approach also implies that we have to adjust the `standard' discount rate for private consumption with an addition to the Simple Ramsey Rule by a substitutability effect ($\eta_{CE_t} g_{E_t}$), to account for how changes in the physical availability of the non-marketed good affect the utility obtained from the private-good. This is unnecessary when using the RPC approach because non-marketed goods are valued in terms of private goods and the RPC effect captures substitutabuility.

To make this concrete and applicable, let us consider the workhorse constant-elasticity-of-substitution (CES) utility function, capturing various degrees of substitutability:
\begin{equation}
\label{eq:utility_CES}
U(C_{t},E_{t}) = \left( \alpha C_{t}^{\frac{\sigma-1}{\sigma}} + (1-\alpha)  \, E_{t}^{\frac{\sigma-1}{\sigma}} \right)^{\frac{\sigma}{\sigma-1}}\ ,
\end{equation}
$0<\sigma<+\infty$, is the constant elasticity of substitution between the two goods, and $0<\alpha<1$ is the utility share parameter for private consumption. The utility function given by equation \ref{eq:utility_CES} is strictly concave, represents homothetic preferences, and both the private good, $C_{t}$, and non-market good, $E_{t}$, are normal. 
It turns out that with CES preferences and imperfect complements, i.e.\ $\sigma>0$, we get the following straightforward equivalence between the dual discounting and RPC  approaches \citep{weikard2005discounting}:
\begin{equation}
RPC_t  \,\,=\,\, \frac{1}{\sigma} \left[  g_{C_t} - g_{E_t}  \right] \,\,= \,\, r_{C_t} - r_{E_t} .
\label{EqRPE}
\end{equation}
Accordingly, the choice of whether one adjusts the numerator via a relative price effect adjustment or the denominator via the use of dual discount rates is not of theoretical importance in intertemporal valuation exercises. In the setting of CES preferences, \cite{ebert2003environmental} has shown that the constant elasticity of substitution between a market good and a non-market good is directly and inversely related to the income elasticity of WTP, $\xi$, of the non-market good \citep[cf.][]{baumgartner2017income}. We can thus write the \textit{RPC} as: 

\begin{equation}
RPC_t  \,\,=\,\, \xi\, \left[  g_{C_t} - g_{E_t}  \right] ,
\label{EqRPE_IEWTP}
\end{equation}
which serves as the key equation our empirical analysis seeks to calibrate.

\bigskip
\section{Empirical strategy}

We build on previous work to estimate income elasticities of WTP for ecosystem services based non-market valuation studies  \citep[e.g.,][]{barrio2010meta, heckenhahn2024relative, jacobsen2009there, richardson2009total, subroy2019worth, }. Our meta-analysis collects mean WTP and mean income estimates at the valuation exercise level, which are then used to estimate income elasticities of WTP and, on their basis, to determine the elasticities of substitution or complementarity between ecosystem services and market goods via their indirect relationship  \citep[cf.][]{baumgartner2015ramsey, ebert2003environmental,  heckenhahn2024relative}. In this section, we first discuss the meta-analysis and subsequently the empirical strategy to derive  estimates of $\xi$. Finally, we discuss the computation of growth rates of ecosystem services.

\medskip
\subsection{Meta-analysis of mean WTP-income value pairs}

The data basis for our analysis is a meta-analysis of existing WTP studies. The main process of dataset creation ran from spring 2022 to early 2023, with additional revisions in summer 2024. In the first phase, we identified potentially relevant non-market valuation studies through a keyword-based search string provided in Appendix~\ref{Search string} in the database SCOPUS. In particular, here, we built on the authors' experience \citep[e.g.,][]{drupp2020between, heckenhahn2024relative, moore2024synthesis} and beta testing. To ensure better comparability of ecosystem service valuation estimates, we focused our search on contingent valuation (CV) studies that were published in peer-reviewed, English-language literature since the year 2000. The keyword-based search resulted in a preliminary data set where each row is a peer-reviewed journal article in which we expect to find relevant (mean) WTP estimates and income data.  Generally, the employed search string was intended to cast a wide net. That is, we expected to later drop several studies due to irrelevance and informational shortcomings. 

The data was then evaluated using the exclusion criteria reported in Appendix~\ref{Exclusion criteria}. After the application of the first exclusion criterion---including whether each article has been cited at least once in SCOPUS---2,174 articles remained. The next exclusion criteria step is an abstract screening to check whether the articles potentially report new, CV-based WTP estimates at all. Strictly theoretical papers as well as reviews, secondary source estimates, and those focused on benefit transfer were excluded to avoid double-counting estimates. Naturally, whether we could access the articles was important but rarely proved to be an issue. At this stage, 1,165 studies remained on which to conduct a detailed screening and subsequent data harvesting.

From the data set of 1,165 WTP studies, we selected a random sample of 100 studies as the basis to fine-tune the screening and coding processes and improve consistency between our two independent coders, we then proceeded to code the full sample. Each paper was carefully scrutinized for appropriate WTP and income data (see Appendix~\ref{Exclusion criteria} for details). A recurring issue was that several papers do not report whether income data is net of taxes or gross income. We have subsequently contacted each paper's corresponding author in search of clarification, with a response rate of around 40 percent.\footnote{If no answer was received, we coded the net-gross dummy as "unclear".} The review of each paper and harvesting of relevant data was a particularly time-intensive process. However, we found it easier to first screen for the inclusion of both mean WTP and mean income estimates---or the information necessary to derive such estimates---before harvesting all relevant data. We also found that there is an important distinction on per-use estimates versus other scales. When per-use estimates have not been paired with the number of uses on a per-respondent scale, they are not comparable---we do not know whether respondents with higher or lower willingness to pay also access the service more or less. As such, per use estimates are set aside. 

Finally, some studies report different scales of ecosystem services, say a 10-percent versus 25-percent versus 50-percent level of provision. When this is the case, we take the most marginal estimate---the smallest change from current conditions. This applies to 69 estimates from 16 studies. The most marginal estimate results from the respondent valuing the level of ecosystem service most similar to the current state---the closest to their lived experience.

Our main analysis builds on studies surviving our exclusion criteria and containing at least the minimum necessary information---a mean WTP estimate and mean respondent income estimate. An unfortunate but necessary result of our focus on comparability is a substantially reduced number of studies contributing to the end result. Of the 1,165 studies passing the first two rounds of screening, {\CountWTPStudies} studies containing {\CountWTPEstimates} distinct WTP-income pairs are of use. Table~\ref{tab:datasummary} provides summary statistics of our sample. Appendix~\ref{AppendixB} includes graphical illustrations of the meta-analysis data. Appendix E, available \href{https://www.dropbox.com/scl/fi/rvt3pu3eaozsmg6a2h9ps/Appendix_E_14-11-24.pdf?rlkey=tc07tfmmv6pbf6cxjl7n946i3&st=ps07oa17&dl=0}{online}, provides the full list of included studies and their respective references.

\begin{table}[t]
\vspace{-0.3cm}
	\centering
	\onehalfspacing
	\caption{Prepared data set description}
		\label{tab:datasummary}
	\begin{tabular}{lll}
		\hline\hline
		Variable	&  Context & Value \\[0.1cm]  \hline
		Countries represented	 & Count & 74 \\[0.1cm] 
            Continent & Observations & \\[0.1cm]
            \hspace{3mm}North America && 88\\[0.1cm]
            \hspace{3mm}South America && 37\\[0.1cm]
            \hspace{3mm}Africa && 35\\[0.1cm]
            \hspace{3mm}Europe && 269\\[0.1cm]
            \hspace{3mm}Asia && 300\\[0.1cm]
            \hspace{3mm}Australia && 6\\[0.1cm]
            Study year & Mean (s.d.) & 2009 (6.5) \\[0.1cm]
            Income & Mean annual, 2020 USD (s.d.) & 38,092  (27,297) \\[0.1cm]
            WTP & Mean annual, 2020 USD (s.d.) & 164 (532) \\[0.1cm]
            Survey sample size & Mean (s.d.) & 665 (851) \\[0.1cm]
            Respondent age & Mean (s.d.) & 43 (6.6) \\[0.1cm]
            Respondent household size & Mean (s.d.) & 3.7 (1.3) \\[0.1cm]
            Forest-relevant estimates & Share of observations & 0.24 \\[0.1cm]
            \hline\hline
	\end{tabular}
	\singlespacing
	\vspace{-0.25cm}
	\begin{minipage}{0.75\textwidth} 
	    \flushleft{\footnotesize \emph{Notes:} s.d. is the standard deviation of the data referenced. Based on N={\CountWTPEstimates} WTP-income pairs contained in {\CountWTPStudies} unique studies.}
   	\end{minipage}
	\vspace{0.3cm}
\end{table}

Further, Table~\ref{tab:datasummary2} presents the number of WTP estimates associated with the regulating and cultural ecosystem services, as introduced in the Millennium Ecosystem Assessment \citep{assessment2005millennium}\footnote {Note that while we linked WTP estimates to all distinct regulating service types presented within the \cite{assessment2005millennium} framework, we focused on only two cultural service types, thereby excluding services like sense of place, laid out in the framework, as well.}, along with estimates related to biodiversity and forest services. Also, for each service type, an example study from our dataset is provided.

 Note that as WTP estimates are often associated with more than one service type, the total number of observations for different ecosystem service types exceeds {\CountWTPEstimates}, with observations roughly evenly distributed between regulating and cultural services.\footnote{For our dataset, WTP estimate overlapping is particularly common between climate and air quality regulation, as well as between aesthetic values and recreation \&  ecotourism.} We control for the number of services a WTP estimates relates to in our analysis.

\begin{minipage}{1.0\textwidth} 
\begin{longtable}[!pt]{>{\small\raggedright\arraybackslash}p{3.55cm} @{\hspace{10pt}} >{\small\raggedright\arraybackslash}p{6.75cm} >{\small\raggedright\arraybackslash}p{0.75cm}  >{\small\raggedright\arraybackslash}p{2.35cm} l l}
\caption{Ecosystem service categories} \label{tab:datasummary2} \\

\hline\hline
Category & Description & N & Example study  \\ \hline

\textbf{Regulating services} &  &  & \\[0.15cm]
Water regulation & Land cover changes influence runoff, flooding, and aquifer recharge. & {\WaterregulationN} & \cite{bliem2012willingness}	 \\[0.1cm]
Air quality regulation & Releasing and absorbing chemicals from the atmosphere. & {\AirqualityregulationN} & \cite{guo2020pollution}\\[0.1cm]
Climate regulation & Local:\ Land cover affects temperature and precipitation; global:\ Sequestration or emission of greenhouse gases. & {\ClimateregulationN}  & \cite{tolunay2015wtp_carbon_sequestration} \\[0.1cm]
Erosion regulation & Vegetative cover is vital for soil retention and landslide prevention. & {\ErosionregulationN}  &  \cite{huenchuleo2012social_psychology}\\[0.1cm]
Water purification \& waste treatment & Ecosystems can be sources of impurities but filter and decompose organic waste and assimilate and detoxify compounds. & {\WaterpurificationregulationN} & \cite{tziakis2009wtp_wastewater_treatment} \\[0.1cm]
Natural hazard regulation & Coastal ecosystems (e.g., mangroves) reduce hurricane and wave damages. & {\NaturalhazardregulationN} & \cite{petrolia2009barrier_islands} \\[0.1cm]
Disease regulation & Ecosystems alter abundance of human pathogens (e.g., cholera) and disease vectors (e.g., mosquitos). &  32 & \cite{Adams2020} \\[0.1cm]
Pest regulation & Ecosystems influence the prevalence of crop and livestock pests and diseases. & 6 & \cite{Adams2020}  \\[0.1cm]
Pollination & Ecosystems influence the distribution, abundance, and effectiveness of pollinators. & 3 & \cite{mwebaze2018bee_pollination} \\[0.9cm]

\textbf{Cultural services} & &  & \\[0.15cm]
Aesthetic values & People enjoy nature's beauty, for instance, in parks and scenic drives. & {\AestheticvaluesN} & \cite{maharana2000valuing_ecotourism_sikkim}\\[0.1cm]
Recreation \& ecotourism & People chose leisure activities based on natural or cultivated landscape traits. & {\RecreationandecotourismN} & \cite{ma2021ecotourism_china}\\[0.1cm]
Spiritual \& religious values & Religions attribute spiritual significance to ecosystems or their components. &  {\SpiritualandreligiousvaluesN} & \cite{endalew2020willingness} \\[0.9cm]
\textbf{Additional sub-types} &  &  & \\[0.15cm]
Biodiversity & Supports ecosystem resilience, productivity, and key functions, enabling regulating and cultural services. & {\BiodiversityN}  & \cite{meyerhoff2012valuing_benefits}\\[0.25cm]
Forest services (non-market) & Provide regulating services, such as climate and air quality regulation, and cultural services. & {\ForestN} & \cite{alassaf2015applying}\\[0.25cm]
\hline\hline
\end{longtable}
\vspace{0.25cm}

\begin{minipage}{0.99\textwidth} 
    \flushleft{\footnotesize \emph{Notes:} Description of ecosystem services types are taken from the Millennium Ecosystem Assessment \citep{assessment2005millennium} and were adapted for the Table.}\\[0.25cm] 
\end{minipage}
\vspace{0.3cm}
\end{minipage}

Beyond that, our inclusive approach to collecting WTP estimates (see Appendix~\ref{Search string}), characterized by an open definition of ecosystem services to maximize sample size, allows for considerable heterogeneity in the specific valuation object, which remains when focusing on particular service types. In particular, many WTP estimates do not directly refer to the specified ecosystem service as defined but rather to related aspects or proxies.
 Climate regulation studies, for instance, are frequently related to forest conservation efforts, like the Table example study of \cite{tolunay2015wtp_carbon_sequestration}. However, they also include studies focusing on general carbon emission reduction  \citep [e.g.,][]{yang2014wtp_co2_mitigation} or, more specifically, on renewable energy provision \citep [e.g.,][]{dogan2019willingness}. While air quality regulation WTP estimates are often linked to forest conservation as well \citep [e.g.,] []{schlapfer2020behavioral_economics}, they frequently focus on direct air quality enhancements as in \cite {guo2020pollution}.  The downside of our inclusive approach is reduced `commodity consistency' \citep[e.g.,][]{bergstrom2006using, moeltner2014cross} as compared to meta-analysis that focus on very specific ecosystem services. We see this not as a bug but as a key purpose of our analysis, as we seek to obtain an aggregate measure of the elasticity across all ecosystem services as well as measures for key ecosystem service sub-types than can be used as reasonable proxies for performing relative price change adjustments to future WTP estimates.
 
\medskip

\subsection{Estimation strategy}

Due to the nature of meta-analyses, of combining data from studies using differing methodologies, estimating the income elasticity of WTP requires careful consideration. We use a log-log specification on mean WTP and mean income values, to directly capture a constant income elasticity of WTP. We account for the structure of our data by clustering standard errors at the study level. 

Several covariates, which may have a direct effect on WTP, would also affect the estimated coefficient of interest if omitted. 
We categorize potential covariates as those related to survey characteristics, income and WTP measures, or ecosystem service. First, survey characteristics include the survey year, the sample size, the method of elicitation (dichotomous choice, open-ended, or other formats), the format used to collect the study data (written questionnaires, oral interviews, or a mix), and a payment vehicle indicator (tax, donation, use charge, free choice, or a mix) which was previously used by \cite{jacobsen2009there}. In addition, we include indicator variables on observations from countries representing 5 percent or more of our dataset---which applies to four countries. Second, income and WTP characteristics reflect whether income is reported as gross or net, per-person or per-household, and the WTP payment terms (whether reoccurring or one-time). Third, ecosystem service characteristics include the types associated with the respondent’s WTP categorized as in the Millennium Ecosystem Assessment \citep{assessment2005millennium} and outlined in Table~\ref{tab:datasummary2} (e.g., climate regulation). This group also includes indicators on the scope of ecosystem services (i.e., the number of services the WTP estimate pertains to) and the spatial scale of the program, project, or policy (local/regional, national, or international) to which the WTP value is linked. Our main model specification is then:

\begin{equation}
\begin{aligned}
ln(WTP_{ij})=\alpha+ \xi\, ln({INC}_{ij}) +  \sum_{k=1}^{n}  \beta_k  x_{ij}
+\epsilon_{i} 
\end{aligned}
\label{full_spec}
\end{equation}

\noindent
where ${\sum_{k=1}^{n}  \beta_k  x_{ij} }$ indicates the inclusion of our list of  ${n}$ preferred covariates. 

As a sensitivity analysis, we also include covariates on respondent age, household size, survey format (written, oral, hybrid), continent, period (pre- versus post-2011) and whether the ecosystem service involved forests. Age and household size, in particular, are less often available and so would reduce sample size while not being particularly relevant. We estimate a large set of alternative models---based variations of our described covariates, resulting in $2^{15}=32,768$ versions around our main log-log specification (see Figure~\ref{fig:spec-chart_Fig5} in the Appendix). These alternatives are based on either including or excluding each covariate in different combinations but keeping the list of ecosystem services indicator variables together---either including or excluding them as a group to avoid estimating a substantially larger set of alternatives for little gain.
We note that study sample size also varies substantially---we find a mean sample size of 665 with a standard deviation of 851. Our preferred weighting approach is to use the square-root of the sample size at the WTP estimate scale. This implies that we put some weight on sample size but avoid the risk that a few particularly large studies drive our result.\footnote{We compare alternative estimators and observation weights in Figures~\ref{fig:appendix_elasticity_models} and~\ref{fig:appendix_elasticity_weights} in the Appendix}

We use a random effects model to capture both within- and between-variance as the source of estimates, following prior work by \cite{jacobsen2009there} and \cite{heckenhahn2024relative}.\footnote{Ideally, we would have multiple similar observations at the country or study level and could rely on a fixed effects model to control for the omission of invariant factors. However, studies often have specific geographic focuses within countries, in other cases involve international issues, and in yet other cases involve less comparable groups (e.g., urban versus rural, visitors versus locals) and so study- and country-level fixed effects would not, in fact, work as intended. Additionally, 272 of {\CountWTPStudies} studies (69 percent) and {\CountWTPEstimates} estimates (37 percent of the dataset) are single observation studies – or singletons. These provide no within-variance and so are in effect dropped during fixed effect model estimation.} 
This choice is supported by a  Hausman Specification Test contrasting fixed and random effects models (while excluding any singleton studies). So, we proceed with a random effects, multivariate model with the square root of sample size weights as our main specification. 

We present alternative model specifications, including univariate random and fixed effects, multivariate fixed effects, OLS, unweighted, and alternative weighted models (e.g., inverse square root of sample size \citep[cf.][]{subroy2019worth}) in the appendix. Separately, we explore heterogeneity in income elasticities. First, we compare study differences across ecosystem service types. Second, we test how income elasticities differ across continents. Third, we test for differences across periods (pre- and post-2011). Fourth, we test for differences based on whether we include the most marginal, average, or all individual estimates, where alternative scales of provision are explored in studies as discussed in the data section. Finally, we explore whether income elasticity estimates differ in income by comparing different segments of the along the income distribution.

\medskip

\subsection{Growth rates}

We assemble growth rates of ecosystem services to obtain a proxy for a global measure of the shift in the relative scarcity of ecosystem services vis-a-vis human-made goods. These estimates extend and update prior work by \cite{baumgartner2015ramsey}, who found that ecosystem services have overall declined by half a percent in the last decades. We focus on non-market (and non-rivalrous) ecosystem services, i.e.\ we do not consider provisioning services but solely capture regulating and cultural services. In a first step, we update the data sources employed by \cite{baumgartner2015ramsey}, notably: Forest cover, Living Planet Index (LPI), and IUCN's Red List Index (RLI). We complement this with two additional measures for regulating services that capture highly salient aspects of environmental quality: air quality regulation and climate regulation. We proxy the former by the negative of changes in PM2.5 emissions, i.e.\ counting reductions in emission as an improvement in air quality. We proxy for the latter with the change in the 2C global mean temperature budget---the upper target of the UN Paris Agreement. Table~\ref{tab:growth} shows the individual components, units of measurement, and data sources. 

\begin{table}[t!]
\vspace{-0.3cm}
	\centering
	\onehalfspacing
	\caption{Components and data sources for estimates of growth rates}
		\label{tab:growth}
	\begin{tabular}{lll}
		\hline\hline
		Component	&  Unit of measurement & Data source \\[0.1cm]  \hline
		Forest area	 & Hectare & WorldBank (2023) \\[0.1cm] 
		Living Planet Index (LPI) & Dimensionless & Zoological Society of London,\\ 
		&& and WWF 2022 \\[0.1cm] 
		Red List Index (RLI) & Various & IUCN RedList (2023),\\
	&&	based on \cite{butchart2010global}	\\[0.1cm]  
		Air quality  &  Micrograms per m$^3$ & WorldBank (2023)	\\ 
		(mean annual PM2.5) &&\\[0.1cm] 
		Climate regulation & Degrees Celsius & NOAA (2023)	\\[0.1cm]  \hline
		GDP per capita & US dollars & WorldBank (2023) \\
		\hline\hline
	\end{tabular}
	\vspace{0.2cm}

\end{table}

Within regulating (forest, LPI, RLI, PM2.5, temperature) and cultural services (forest, LPI, RLI) as well as aggregate ecosystem services we take the arithmetic mean of individual components.  To calculate growth rates, we use the time span with the longest comparable data (1993 to 2016) and estimate exponential growth rates, including standard errors. We use the largest standard error of the individual growth rate components---climate for regulating and aggregate services, and the living planet index for cultural services---when aggregating standard errors. Akin to estimating growth rates of ecosystem services, we also estimate the growth rate of global GDP per capita and its standard error. In contrast to \cite{baumgartner2015ramsey}, we do not subtract provisioning services, as we do not examine it as a separate ecosystem service category.\footnote{All time series show a clear trend except for air quality, which deteriorates from 1990 to 2010 and improves again thereafter. We thus also redo the analysis of growth rates for the time frame  2010 to 2016.}

\bigskip
\section{Results}
We now present here estimates of income elasticities of WTP, $\xi$, for ecosystem services globally as well as select regions. We also estimate income elasticities based on subcategories of ecosystem services as well as different time frames. We subsequently couple the estimates of income elasticities with estimates of good-specific growth rates to compute relative price changes of ecosystem services.

\medskip
\subsection{Income elasticity of WTP for ecosystem services}
We first estimate the income elasticity of WTP for aggregate ecosystem services on our full sample with key controls. Our central estimate of the income elasticity of WTP amounts to \num[round-mode=places,round-precision=2]{\MainEstimate} (95-CI: \num[round-mode=places,round-precision=2]{\MainEstimateLB} to \num[round-mode=places,round-precision=2]{\MainEstimateUB}) based on a random effects model.\footnote{By contrast, a quasi-univariate regression -- we still include indicator variables for observations from countries representing 5\% or more of the data -- yields an estimate of the income elasticity of WTP for ecosystem services of \num[round-mode=places,round-precision=2]{\UnivariateEstimate} (95-CI: \num[round-mode=places,round-precision=2]{\UnivariateEstimateLB} to \num[round-mode=places,round-precision=2]{\UnivariateEstimateUB}). Instead of taking the most-marginal estimates when multiple levels of an ecosystem service are described, we could estimate on the average across all levels or treat each level-estimate separately (this applies to 16 studies with 69 total estimates). Estimating instead on the average in these cases results in a nearly identical estimate. If instead we include all individual estimates in these cases, the number of observations rises to {\CountAllSeparateBasisEstimate} and the coefficient estimate on the income elasticity of WTP increases to {\AllSeparateBasisEstimate}. We also note that the inclusion of a single negative WTP estimate changes the main income elasticity of WTP estimate from {\OnlyPositiveWTPEstimates} to \num[round-mode=places,round-precision=2]{\MainEstimate}} We also develop a specification graph to investigate the sensitivity of our estimate to various combinations of control variables which we report in Figure~\ref{fig:spec-chart_Fig5} of Appendix~D. Compared to alternative specifications, our main estimate falls at the {\SpecChartMainEstimatePercentile} percentile. Our main estimate maps into a mean value for the elasticity of substitutability between ecosystem services and market goods of \num[round-mode=places,round-precision=2]{\ElasticityOfSubstitutability} (95-CI: \num[round-mode=places,round-precision=2]{\ElasticityOfSubstitutabilityLB} to \num[round-mode=places,round-precision=2]{\ElasticityOfSubstitutabilityUB}), indicating that ecosystem services and market goods are substitutes.

\begin{table}[h!]
\vspace{0.3cm}
	\centering
	\onehalfspacing
		\caption{Income elasticity of WTP for aggregate ecosystem services}
			\label{tab_IEWTP}
	\begin{tabular}{llll}
		\hline\hline
		ln(INCOME)\quad\quad\quad  & S.E.\quad\quad\quad &  N\quad\quad\quad &  Overall R$^2$ \\ \hline
        \num[round-mode=places,round-precision=2]{\MainEstimate}***& \num[round-mode=places,round-precision=2]{\MainEstimateSE}& {\CountWTPEstimates} & \quad0.28  \\
		\hline\hline
	\end{tabular}
	\singlespacing
	\vspace{-0.25cm}
	\begin{minipage}{0.495\textwidth} 
	    \flushleft{\footnotesize \emph{Notes:} 
     Multivariate regression based on Equation~\ref{full_spec}. Significance levels: * p$<$0.1, ** p$<$0.05, *** p$<$0.01} 
   	\end{minipage}
\vspace{0.4cm}
\end{table}

\begin{figure}[t]
	\caption{Estimates of the income elasticity of WTP for select models and service types.}
	\label{fig:results-with-CIs}
	\centering
	\begin{minipage}{1\textwidth} 
	    \includegraphics[width=1\linewidth]{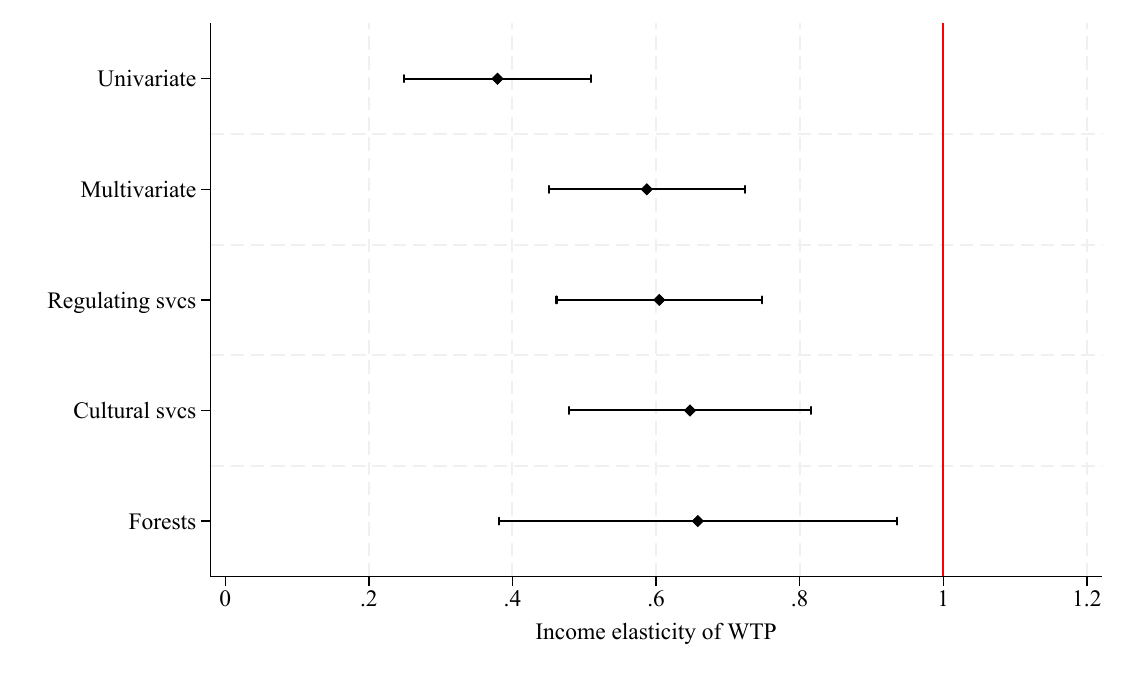 }
	\end{minipage}
 	\begin{minipage}{0.87\textwidth} 
   \flushleft{\footnotesize \emph{Notes:} Estimates are the coefficients on $ln(INCOME)$ from the main and univariate specifications in Table~\ref{tab_IEWTP} as well as estimates based on subsets of observations on regulating services, and cultural services, and forests using the main model. 95 percent confidence interval estimates are included around the point estimates. }
   	\end{minipage}
   	\vspace{0.4cm}
\end{figure}

Estimates on subsets allow us to investigate the extent of heterogeneities. We consider different sub-types of ecosystem services, and potential differences across continents and time frames. Table~\ref{tab_IEWTP_hetES} reports income elasticities of WTP across different sub-types of ecosystem services: regulating and cultural services as well as key sub-categories. We find little variation in income elasticities, noting that oftentimes projects valued in CV studies encompass contributions to multiple services. We find the estimates of the income elasticity for air and water regulation to be lower. We also split the sample into forest and non-forest ecosystem services, as this serves as a key input to our application on natural capital accounting in the \textit{CWON} example in Section~\ref{sec_CWON}. We find that the income elasticity of forest ecosystem services is slightly higher than the aggregate estimate, but far from significantly so. While the 95 CI for our main aggregate estimate does not include the Cobb-Douglas case ($\xi=1$), the 95 CI for non-market forest ecosystem services overlaps into the complements domain. For comparison, we present the univariate and choice set of control estimates alongside key subgroups in Figure~\ref{fig:results-with-CIs}. 

\begin{table}[t!]
\vspace{0.8cm}

	\centering
	\onehalfspacing
	\caption{Heterogeneity of income elasticities of WTP across ecosystem service types}
				\label{tab_IEWTP_hetES}
	\begin{tabular}{lllll}
		\hline\hline
		&  
		ln(INCOME)\quad\quad  & S.E.\quad\quad\quad &  N\quad\quad\quad &  Overall\ R$^2$  \\ \hline
  \textbf{Regulating Services}             & \num[round-mode=places,round-precision=2]{\RegulatingServicesEst}***  & \num[round-mode=places,round-precision=2]{\RegulatingServicesSE}& {\RegulatingServicesN} & {\RegulatingServicesRsquared}  \\ \hline
        Water regulation 			    & \num[round-mode=places,round-precision=2]{\WaterregulationEst}***  & \num[round-mode=places,round-precision=2]{\WaterregulationSE}& {\WaterregulationN} & {\WaterregulationRsquared} \\
		Air quality regulation 		    & \num[round-mode=places,round-precision=2]{\AirqualityregulationEst}***  & \num[round-mode=places,round-precision=2]{\AirqualityregulationSE}& {\AirqualityregulationN} & {\AirqualityregulationRsquared} \\
        Climate regulation 			    & \num[round-mode=places,round-precision=2]{\ClimateregulationEst}***  & \num[round-mode=places,round-precision=2]{\ClimateregulationSE}& {\ClimateregulationN} & {\ClimateregulationRsquared}  \\
        Erosion regulation              & \num[round-mode=places,round-precision=2]{\ErosionregulationEst}***  & \num[round-mode=places,round-precision=2]{\ErosionregulationSE}& {\ErosionregulationN} & {\ErosionregulationRsquared} \\ 
        Water purification \& waste treatment & \num[round-mode=places,round-precision=2]{\WaterpurificationregulationEst}** & \num[round-mode=places,round-precision=2]{\WaterpurificationregulationSE} &{\WaterpurificationregulationN} & {\WaterpurificationregulationRsquared} \\
        Natural hazard regulation       & \num[round-mode=places,round-precision=2]{\NaturalhazardregulationEst}*** & \num[round-mode=places,round-precision=2]{\NaturalhazardregulationSE}&{\NaturalhazardregulationN} & {\NaturalhazardregulationRsquared} \\ \hline
        \textbf{Cultural Services }	            & \num[round-mode=places,round-precision=2]{\CulturalServicesEst}***  & \num[round-mode=places,round-precision=2]{\CulturalServicesSE}& {\CulturalServicesN} & {\CulturalServicesRsquared}  \\ \hline
		Aesthetic values 			    & \num[round-mode=places,round-precision=2]{\AestheticvaluesEst}***  & \num[round-mode=places,round-precision=2]{\AestheticvaluesSE}& {\AestheticvaluesN} & {\AestheticvaluesRsquared}  \\        
		Recreation \& ecotourism 		& \num[round-mode=places,round-precision=2]{\RecreationandecotourismEst}***  & \num[round-mode=places,round-precision=2]{\RecreationandecotourismSE}& {\RecreationandecotourismN} & {\RecreationandecotourismRsquared}  \\
		Spiritual \& religious values	& \num[round-mode=places,round-precision=2]{\SpiritualandreligiousvaluesEst}***  & \num[round-mode=places,round-precision=2]{\SpiritualandreligiousvaluesSE}& {\SpiritualandreligiousvaluesN} & {\SpiritualandreligiousvaluesRsquared}  \\ \hline
		        \textbf{Additional sub-types}	            &   &  &   \\ \hline
		Biodiversity 	                & \num[round-mode=places,round-precision=2]{\BiodiversityEst}***  & \num[round-mode=places,round-precision=2]{\BiodiversitySE}& {\BiodiversityN} & {\BiodiversityRsquared}  \\
		Forest ecosystem services 	    & \num[round-mode=places,round-precision=2]{\ForestEst}***  & \num[round-mode=places,round-precision=2]{\ForestSE}& {\ForestN} & {\ForestRsquared}  \\ \hline
	\end{tabular}
	\singlespacing
	\vspace{-0.25cm}
	\begin{minipage}{0.85\textwidth} 
	    \flushleft{\footnotesize \emph{Notes:} Multivariate regressions. Too few observations to compute R-squared in one case using the selected regression model. Significance levels: * p$<$0.1, ** p$<$0.05, *** p$<$0.01}
   	\end{minipage}
\vspace{0.4cm}
\end{table}

We next divide our sample by the continent on which the CV study has been undertaken, and report the results in Table~\ref{tab_IEWTP_continents}. We note that the estimates are mostly concentrated in Asia, followed by Europe, with substantially fewer estimates from other regions.\footnote{Several studies from Africa involve day trips and other per-use scenarios and are excluded here.} In terms of income elasticities, we calculate an insignificant estimate for North America, South America, and, surprisingly, Asia while values in Europe and Africa are larger than our main estimate. We find, in general, that continents are too diverse to place much value in continental averages and less relevant than having a global average as a starting point.

\begin{table}[h!]
\vspace{0.5cm}

	\centering
	\onehalfspacing
	
	\caption{Heterogeneity of income elasticities of WTP across continents }
		\label{tab_IEWTP_continents}

	\begin{tabular}{lllll}
		\hline\hline
		&  
		ln(INCOME)\quad\quad  & S.E.\quad\quad\quad &  N\quad\quad\quad &  Overall\ R$^2$ \\ \hline
		North America 	\quad	& {\NorthAmericaEst}  & {\NorthAmericaSE}& {\NorthAmericaN} & {\NorthAmericaRsquared}  \\
		South America		& {\SouthAmericaEst}  & {\SouthAmericaSE}& {\SouthAmericaN} & {\SouthAmericaRsquared}  \\
		Africa 				& {\AfricaEst}***  & {\AfricaSE}& {\AfricaN} & {\AfricaRsquared}  \\
		Europe 				& {\EuropeEst}***  & {\EuropeSE}& {\EuropeN} & {\EuropeRsquared}  \\
		Asia 				& {\AsiaEst}***  & {\AsiaSE}& {\AsiaN} & {\AsiaRsquared}  \\
		\hline\hline
	\end{tabular}
	\singlespacing
	\vspace{-0.25cm}
	\begin{minipage}{0.805\textwidth} 
	    \flushleft{\footnotesize \emph{Notes:} Multivariate regressions. Too few studies in South America and Africa with multiple estimates to base an overall R-squared value on with the selected regression model (insufficient within variance). Significance levels: * p$<$0.1, ** p$<$0.05, *** p$<$0.01}
   	\end{minipage}
\vspace{0.5cm}

\end{table}

The largest prior comparable meta-analysis on the income elasticity of WTP (for biodiversity conservation only) was conducted by \cite{jacobsen2009there}. Their main result was an income elasticity of WTP estimate of 0.38, but published more than a decade ago. It is, thus, interesting to investigate how our estimate of the income elasticity of WTP relates in a more comparable time frame and in comparison to the most recent decade. In Table~\ref{tab_IEWTP_pubyear} we break down the sample by sampling year. We conduct this analysis based on our multivariate estimation strategy. First, we consider estimates from publications based on samples collected up to and including the year 2010 and find an income elasticity of {\FirstPeriodEst} in our full model with controls. In contrast, the income elasticity for 2011 onwards is lower, at {\SecondPeriodEst} (see Table \ref{tab_IEWTP_pubyear}). Thus, overall, the evidence regarding elasticity changes over time is mixed: while based on our data, we find larger elasticity values than \cite{jacobsen2009there}, the income elasticity is significantly lower for the later period within our split sample analysis.\footnote{Two other differences in the meta-analysis by \cite{jacobsen2009there} and ours concern the ecosystem service type under consideration (biodiversity in their case) and whether also grey-literature was included in the analysis. On the first, we do not find evidence that income elasticities are different for biodiversity-related CV studies (see Table~\ref{tab_IEWTP_hetES}). On the latter, \cite{heckenhahn2024relative} find a larger income elasticity estimate when focusing on peer-reviewed literature only in their German case study.}

\begin{table}[!t]
	\centering
	\onehalfspacing
	\caption{Heterogeneity of income elasticities of WTP across decades}
	\label{tab_IEWTP_pubyear}
\begin{tabular}{lllll} 
	\hline\hline
	&  
	ln(INCOME)\quad\quad  & S.E.\quad\quad\quad &  N\quad\quad\quad &  Overall\ R$^2$ \\ \hline
	pre-2011 		& {\FirstPeriodEst}***  & {\FirstPeriodSE}& {\FirstPeriodN} & {\FirstPeriodRsquared}  \\
	2011-2021		& {\SecondPeriodEst}***  & {\SecondPeriodSE}& {\SecondPeriodN} & {\SecondPeriodRsquared}  \\ 
	\hline\hline
\end{tabular}
	\singlespacing
	\vspace{-0.25cm}
	\begin{minipage}{0.85\textwidth} 
	    \flushleft{\footnotesize \emph{Notes:} Multivariate regression-based. Excludes studies where the year of data collection is uncertain and the study authors could not be contacted for clarification. Significance levels: * p$<$0.1, ** p$<$0.05, *** p$<$0.01}
   	\end{minipage}
\vspace{0.4cm}
\end{table}

Finally, we examine whether income elasticity estimates differ across income levels. Previous work by \cite{barbier2017income} and \cite{ready2002relationship} had suggested that estimates of income elasticities might increase along income levels by examining data in primary CV studies. Here, we now test how estimates of income elasticity differ across income levels in our aggregate-level data set.\footnote{Note there could be differences in the types of ecosystem services valued across high- and low-income countries that could introduce omitted variables bias in our meta-regression analyses. For instance, regarding fishing activities, in high-income countries, WTP might rather be associated with the recreational domain, while in low-income countries, even though we generally exclude WTP explicitly focusing on provisioning services, people may tend to perceive fishing as more of a means of subsistence.} 
To this end, we first consider a median split and find comparable estimates when groups as below versus above (inclusive) median income. We explored further ways of cutting the data, using thirds, quartiles and quintiles as well. For instance, when comparing the bottom with the top quartiles. We find a substantially larger income elasticity of WTP for the top 25 percent income group. Overall, we thus find some evidence that the income elasticity of WTP may increase along income levels. 

\begin{table}[h!]
\vspace{0.4cm}
	\centering
	\onehalfspacing
		\caption{Income elasticity of WTP for  ecosystem services across income brackets}
			\label{tab_IEWTP}
	\begin{tabular}{lllll}
		\hline\hline
		Sample	&  
		ln(INCOME)\quad\quad\quad  & S.E.\quad\quad\quad &  N\quad\quad\quad &  Overall R$^2$ \\ \hline
        Below median        & {\BelowMedianEst}***& {\BelowMedianSE}& {\BelowMedianN} &\quad {\BelowMedianRsquared}  \\
        Above median        & {\AboveMedianEst}***& {\AboveMedianSE}& {\AboveMedianN} & \quad {\AboveMedianRsquared}  \\ \hline
       Bottom 25\%          & {\BottomQuarterEst}***& {\BottomQuarterSE}& {\BottomQuarterN} &\quad {\BottomQuarterRsquared}  \\
        Top 25\%            & {\TopQuarterEst}***& {\TopQuarterSE}& {\TopQuarterN} & \quad {\TopQuarterRsquared}  \\
		\hline\hline
	\end{tabular}
	\singlespacing
	\vspace{-0.25cm}
	\begin{minipage}{0.8125\textwidth} 
	    \flushleft{\footnotesize \emph{Notes:} Multivariate regressions. The set of controls including the study year, sample size, income information (gross/net, individual/household), payment type and elicitation method. Significance levels: * p$<$0.1, ** p$<$0.05, *** p$<$0.01}
   	\end{minipage}
\vspace{0.4cm}
\end{table}

\medskip

\subsection{Growth rates}

Table~\ref{tab_growth} reports estimates on the growth rates of ecosystem service categories and their standard errors, alongside the growth rate of GDP per capita. Growth metrics are estimated based on data for the longest common time frame, for the years 1993 to 2016. 

\begin{table}[h!]
\vspace{0.4cm}
	\centering
	\onehalfspacing
	\caption{Good-Specific Growth Rates}
		\label{tab_growth}
	\begin{tabular}{ll}
		\hline\hline
		Indicator	& Annual growth rate (S.E.) \quad\quad\quad  \\ \hline
		Forest area	 & -0.11\%  (0.00\%) \\ 
		Living planet index & -2.84\% (0.06\%) \\
		Red list index & -0.42\% (0.01\%)	\\ 
		Air quality (PM2.5) & -0.16\% (0.17\%)	\\ 
		Climate regulation & -1.50\% (0.14\%)	\\ \hline
		Aggregate Ecosystem Services\quad \quad & -1.01\% (0.17\%) \\  \hline
		GDP per capita& \ 1.82\% (0.02\%) \\
		\hline\hline
	\end{tabular}
	\vspace{0.4cm}
\end{table}

We find substantial heterogeneity in growth rates. The Living Planet Index and climate regulation metrics show the largest negative rates, while the change in forest area and air quality metrics show the lowest rates of change.\footnote{Results are qualitatively similar when constraining the analysis to the most recent trend data, except for air quality regulation which shows a positive development in the current trend data (2010 to 2016), improving by 1.78\% per year. In contrast, the decline rate for climate regulation is more strongly negative. Overall, we find a somewhat smaller rate of de-growth of -0.73 percent for the time period 2010 to 2016.} Our estimate of aggregate ecosystem service change is -1.01 percent (CI: -1.34 to -0.68), while GDP per capita has increased by 1.82 percent (CI: 1.78 to 1.86) over the same period. This amounts to a sizable shift in the relative scarcity of ecosystem services vis-a-vis market goods. Ecosystem services have thus become relatively scarcer by 2.83 percent per year.

\medskip

\subsection{Relative price changes of ecosystem services}

We can now combine the two critical pieces---the income elasticity and growth rate estimates---to compute relative price changes (\textit{RPC}). Table~\ref{tab_RPC} reports our estimates of \textit{RPCs} both in the aggregate and for different ecosystem service categories.

\begin{table}[h!]
	\centering
	\onehalfspacing
	\caption{Relative price changes (RPC) of ecosystem services}
		\label{tab_RPC}
	\begin{tabular}{llll}
		\hline\hline
		Sample\quad\quad 	&  
		$ \xi=1/\sigma$ (S.E.) \quad\quad & $g_C-g_E$ (S.E.) \quad\quad\ &  $RPC $ (C.I.) \\[0.1cm] \hline
		Regulating Services \quad\quad & \num[round-mode=places,round-precision=2]{\RegulatingServicesEst} (\num[round-mode=places,round-precision=2]{\RegulatingServicesSE}) &{\GrowthDiffGDPcapitaRegSvcsEst}\% ({\GrowthDiffGDPcapitaRegSvcsSE}\%) & \num[round-mode=places,round-precision=2]{\RPCRegulatingServices}\% \\&&&(\num[round-mode=places,round-precision=2]{\RPCRegulatingServicesLB}\% to \num[round-mode=places,round-precision=2]{\RPCRegulatingServicesUB}\%)   \\[0.1cm]  
		Cultural Services 					& \num[round-mode=places,round-precision=2]{\CulturalServicesEst} (\num[round-mode=places,round-precision=2]{\CulturalServicesSE}) &{\GrowthDiffGDPcapitaCulSvcsEst}\% ({\GrowthDiffGDPcapitaCulSvcsSE}\%) & \num[round-mode=places,round-precision=2]{\RPCCulturalServices}\% \\&&&(\num[round-mode=places,round-precision=2]{\RPCCulturalServicesLB}\% to \num[round-mode=places,round-precision=2]{\RPCCulturalServicesUB}\%)   \\[0.1cm] \hline
		Aggregate Services & \num[round-mode=places,round-precision=2]{\MainEstimate} (\num[round-mode=places,round-precision=2]{\MainEstimateSE}) &{\GrowthDiffGDPcapitaAggEst}\% ({\GrowthDiffGDPcapitaAggSE}\%) & \num[round-mode=places,round-precision=2]{\RPCAggregateServices}\%  \\&&&(\num[round-mode=places,round-precision=2]{\RPCAggregateServicesLB}\% to \num[round-mode=places,round-precision=2]{\RPCAggregateServicesUB}\%) \\[0.1cm]  \hline
		Forest Services & \num[round-mode=places,round-precision=2]{\ForestEst} (\num[round-mode=places,round-precision=2]{\ForestSE}) &{\GrowthDiffGDPcapitaForestsEst}\% ({\GrowthDiffGDPcapitaForestsSE}\%) & \num[round-mode=places,round-precision=2]{\RPCForestServices}\% \\&&&(\num[round-mode=places,round-precision=2]{\RPCForestServicesLB}\% to \num[round-mode=places,round-precision=2]{\RPCForestServicesUB}\%)\\[0.1cm] 
		\hline\hline
	\end{tabular}
	\vspace{0.35cm}
 	\begin{minipage}{1.5\textwidth} 
	    \flushleft{\footnotesize \emph{Notes:} RPC 95\% confidence interval estimates based on ${\xi(g_C-g_E)} \pm \text{1.96} \times \sqrt{\left(\frac{S.E.(\xi)}{\xi}\right)^2 + \left(\frac{S.E. (g_C-g_E)}{g_C-g_E}\right)^2}$.
     }
   	\end{minipage}
\vspace{0.2cm}
\end{table}

Our central estimate for the \textit{RPC} of aggregate ecosystem services is \num[round-mode=places,round-precision=2]{\RPCAggregateServices} percent (CI: \num[round-mode=places,round-precision=2]{\RPCAggregateServicesLB} to \num[round-mode=places,round-precision=2]{\RPCAggregateServicesUB}). 
That is, the value of ecosystem services is increasing by around \num[round-mode=places,round-precision=1]{\RPCAggregateServices} percent per year relative to market goods. This is substantially larger than the estimate reported in \cite{baumgartner2015ramsey}. The \textit{RPC} estimate for regulating services is only slightly higher than that for cultural services, which is qualitatively similar to what \cite{heckenhahn2024relative} find for a German case study. While the income elasticity for forest ecosystem services is higher than for ecosystem services on aggregate, the rate of decline of forest area is considerably smaller; in combination, the \textit{RPC} of forest ecosystem services (\num[round-mode=places,round-precision=2]{\RPCForestServices} percent) is smaller than that of aggregate ecosystem services.

\bigskip
\section{Application to Environmental-Economic Accounting}
\label{sec_CWON}
  
Relative price adjustments of ecosystem services are relevant for both policy appraisal and environmental-economic accounting. Here, we explore implications for accounting, considering the \textit{CWON} 2021 report by the \cite{world2021changing} as a prominent case to illustrate the approach and its importance with a focus on forest natural capital.\footnote{We have subsequently also applied the approach to proposing adjustments for assessing changes to ecosystem services in benefit-cost analysis \citep{drupp2024accounting}} We afterwards illustrate implications also for our aggregate measure of ecosystem services.
 
\textit{CWON}, like most measures of comprehensive wealth, only features selected natural capital stocks, predominantly relating to fossil energy resources and other provisioning services that are traded on markets. \textit{CWON}, however, also considers non-timber forest benefits as part of its natural capital accounting. Non-timber forest benefits are currently estimated to be around 12 percent of the total value of natural capital \citep{world2021changing}. Non-timber ecosystem service values in the year 2018, in WTP per hectare, were based on a meta-regression analysis drawing on 270 estimates from non-market valuation studies of non-timber forest benefits by \citep{siikamaki2021global}. Per-hectare values are assumed to be constant over time and only adjusted for inflation by using country-specific GDP deflators \citep{world2021changing}.  The capitalized value of non-timber ecosystem services is calculated as the present value of annual services, discounted over a 100-year time horizon at a constant discount rate of 4 percent. This implies that no adjustment for \textit{RPCs} is factored in despite forest de-growth, particularly in comparison to GDP per capita. Implicitly, this carries the assumption that WTP does not increase with income and---in the setting of our model---that ecosystem services are considered perfect substitutes to market goods.\footnote{\cite{siikamaki2021global} report positive and significant GDP elasticities of WTP for recreation and habitat/species conservation, for instance, but these are not considered in the \textit{CWON} natural capital valuation.} 

\begin{figure}[!pth]
            	\caption{Accounting for public forest natural capital with changing relative prices.}
\label{fig:main-result-CWON}
  \centering
      	    \subfloat[Normalized projected real growth]
	    	{\includegraphics[width=.49\textwidth]{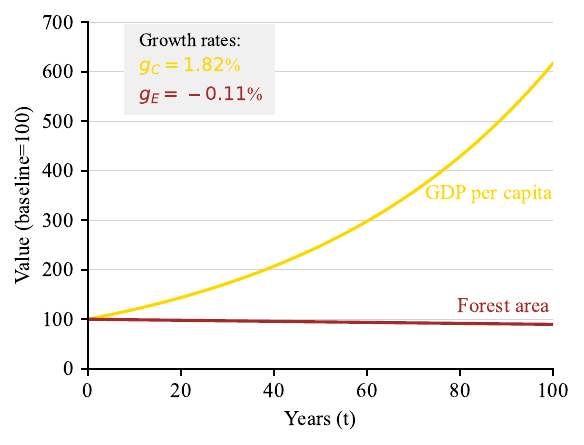}}\label{fig:A}
	    	 \subfloat[Yearly relative price change (\textit{RPC})]
	    	{\includegraphics[width=.49\textwidth]{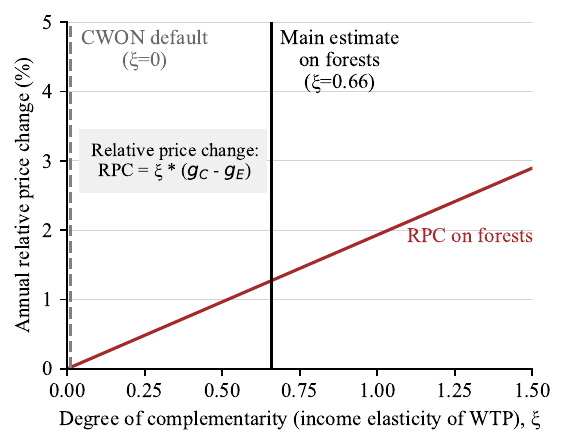}}\label{fig:quadrpr}	\\[0.21cm]
	    		    \subfloat[Evolution of WTP]
	    	{\includegraphics[width=.49\textwidth]{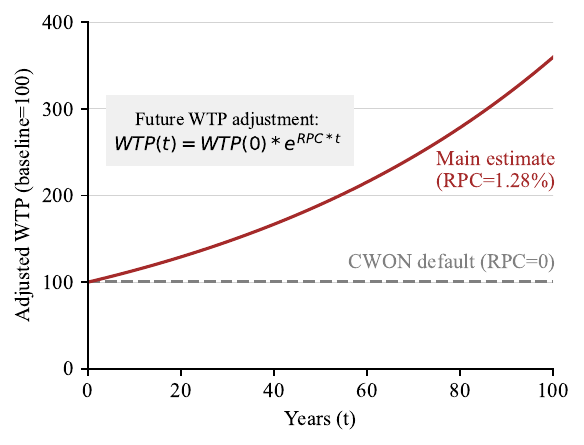}}\label{fig:pm25local}
      	    \subfloat[Increase in PV of forest natural capital]
	    	{\includegraphics[width=.49\textwidth]{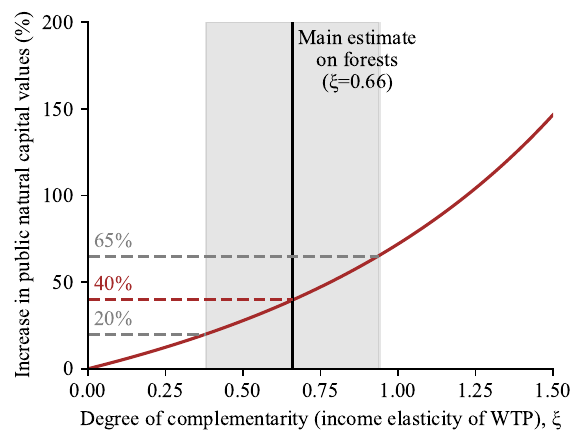}}\label{fig:pm25local}\\[0.1cm]
       	\begin{minipage}{0.97\textwidth} 
   \flushleft{\footnotesize \emph{Notes:} Panel (a): Relative to growth in market goods (or real income, reflected by GDP per capita), global forest area has been decreasing, which we here project forward. Initial values are normalized to 100 in year 0. Panel (b): The relative price change (\textit{RPC}) rule maps growth rates of GDP per capita and of ecosystem services into yearly relative price adjustments against the rate at which WTP for ecosystem services changes with income. Panel (c): Future WTP adjustment when applying our main estimate for the RPC for forest ecosystem services.  Panel (d) shows the estimated increase in The Changing Wealth of Nations' (\textit{CWON})  non-timber forest natural capital value (in \%), relative to the \textit{CWON}'s current estimate, as a function of the degree of complementarity between forest ecosystem services and market goods, measured by the income elasticity of WTP (see the maroon line). The vertical black line indicates the central estimate of the income elasticity of WTP for forest ecosystem services, while the grey-shaded area indicates its 95 confidence interval. Horizontal, dashed helplines indicate the corresponding increase in the public natural capital values (in \%). }
   	\end{minipage}
   	\vspace{0.35cm}
\end{figure}

 Taking our estimated growth rates for forest area and for GDP per-capita as best estimates of growth rates for the 100 year time horizon in question (see Panel (a) in Figure~\ref{fig:main-result-CWON}), we compute \textit{RPCs} for forest ecosystem services using our disentangled estimate on the income elasticities of WTP for forest ecosystem services (see Panel (b) in Figure~\ref{fig:main-result-CWON}). We use the \textit{RPC} of \num[round-mode=places,round-precision=2]{\RPCForestServices} percent to adjust future WTP estimates for increasing income and changing real scarcities of forest ecosystem services, and contrast these yearly adjusted WTPs with the \textit{CWON} default which considers constant real WTPs over the time horizon (see Panel (c) in Figure~\ref{fig:main-result-CWON}). Real WTP in 30 (100) years, for instance, would be {\UpliftingForestsAtThirty} ({\UpliftingForestsAtOneHundred}) percent higher as compared to the current \textit{CWON}, which does not consider relative price changes. 
 
 We then compute the discounted present value of non-timber forest natural capital, using  \textit{CWON}'s discount rate of 4 percent, and compare it to the unadjusted value from \textit{CWON}. In Panel (d) of Figure~\ref{fig:main-result-CWON}, we 
depict the estimated increase in the non-timber forest natural capital value (in \%), relative to the \textit{CWON}'s current estimate, as a function of the degree of complementarity between forest ecosystem services and market goods, measured by the income elasticity of WTP for forest ecosystem services. For our central estimate of the \textit{RPC} of forest ecosystem services, we find that the value of non-timber forest natural capital should be uplifted by {\ForestsUpliftingMed} percent, with a 95 percentile confidence interval around the income elasticity resulting in a range of uplift-factors of {\ForestsUpliftingLB} to {\ForestsUpliftingUB} percent (see Panel (d) in Figure~\ref{fig:main-result-CWON}). Alternatively, Cobb-Douglas substitutability ($\sigma=\xi=1$) would imply uplifting the present value of non-timber forest ecosystem services by 72 percent. Another prominent assumption in applied modelling is to use an elasticity of substitution of 0.5 \citep[c.f.,][]{sterner2008even}, i.e., an income elasticity of 2 (off the chart here), would translate into uplifting the public natural capital value by around 280 percent. 

Considering the limited degree of substitutability and shifts in relative scarcity by performing \textit{RPC} adjustments in computing the natural capital value of non-timber forest services makes a material difference to natural capital accounting in \textit{CWON}. The {\ForestsUpliftingMed} percent increase in non-timber forest value would lead to an increase of the overall natural capital value in \textit{CWON} of around 5 percent. 

\begin{figure}[t!]
	\caption{Increase in public natural capital values along the degree of complementarity.}
   	\label{fig:main-results-aggregate}
	\centering
	\begin{minipage}{0.9\textwidth} 
	    \includegraphics[width=1\linewidth]{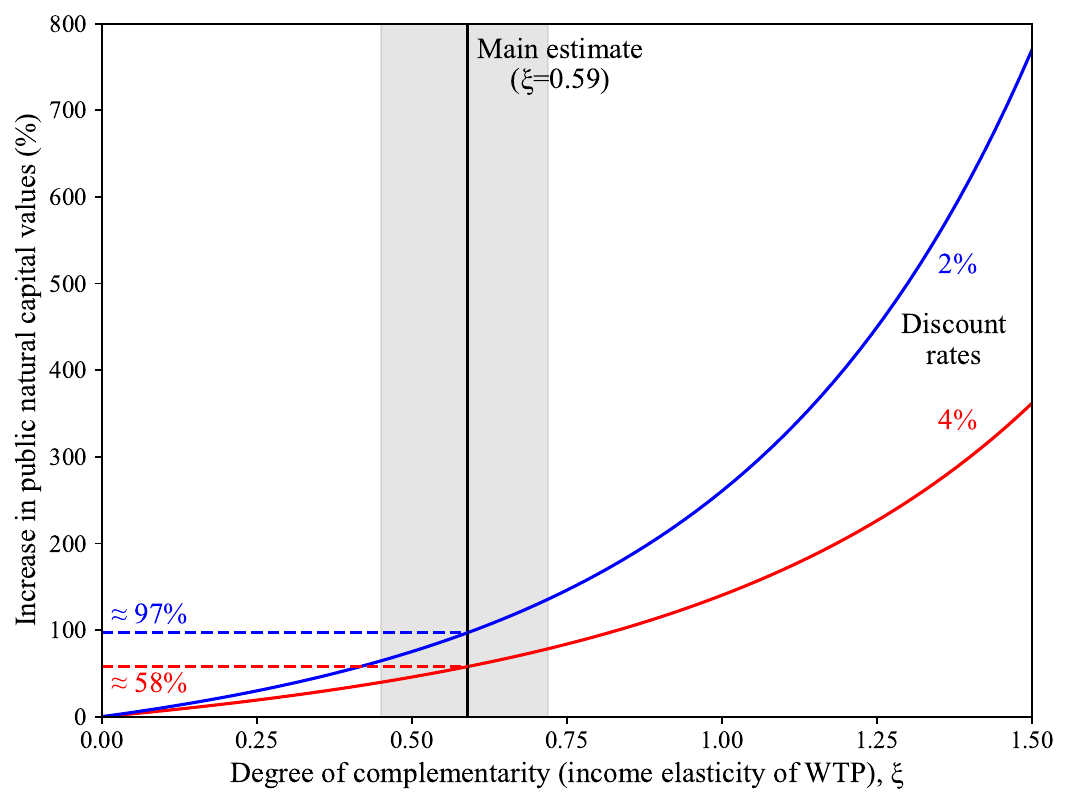}
	\end{minipage}
 	\begin{minipage}{0.95\textwidth} 
   \flushleft{\footnotesize \emph{Notes:} Estimated increase in public natural capital values for our aggregate assessment of ecosystem services (in \%), relative to a case where relative price changes are not considered, as a function of the degree of complementarity between ecosystem services and market goods, measured by the income elasticity of WTP for ecosystem services. 
   The red and blue lines illustrate effects for different discount rates of 4\% (red, as in CWON guidance) and 2\% (blue, as in US Circular A-4). 
   The vertical black line indicates the central estimate of the income elasticity of WTP for our aggregate assessment of ecosystem services, while the grey-shaded area indicates its 95  confidence interval. Horizontal, dashed helplines indicate the corresponding increase in the public natural capital value (in \%).}
   	\end{minipage}
   	\vspace{0.5cm}
\end{figure}

  Beyond the \textit{CWON} case study, we illustrate implications also for our aggregate measure of ecosystem services. Using, the \textit{RPC} of aggregate ecosystem services, which draws on a slightly lower income elasticity of WTP but a larger difference in growth rates, due to a stronger decline in aggregate ecosystem services, we obtain a central uplift-factor for public natural capital of {\AggregateUpliftingLB} percent (see Figure~\ref{fig:main-results-aggregate}), which amounts to a \num[round-mode=places,round-precision=0]{\UpliftingDifference} percent increase as compared to the \textit{CWON} uplift factor. When changing the discount rate from  \textit{CWON}'s 4 percent to a rate of 2 percent, as per current guidance in US Circular A-4 and as recommended by most experts \citep{drupp2018discounting}, we find that the public natural capital value should be uplifted by around {\AggregateUpliftingLB} percent according to our main estimate for the income elasticity of WTP (see Figure~\ref{fig:main-results-aggregate}). 

\medskip
\medskip

\newpage
\section{Discussion}

Estimating the trajectory of shadow prices for ecosystem services requires a theoretical structure in order to project into the future. Furthermore, the use of the income elasticity of WTP as a proxy for the degree of limited substitutability rests on particular assumptions regarding social preferences. On the empirical side, our study identifies the degree of complementarity via the income elasticity of WTP for ecosystem services based on a meta-analysis of the peer-reviewed literature. {\CountWTPEstimates} unique (mean) income-WTP pairs are considered across studies and geographical contexts over a 20 year time frame. Given these data,  assumptions are also required to allow the aggregation of ecosystem services and the computation of ecosystem growth rates across the study samples. The assumptions are discussed here. We argue that our estimates of relative price increases could well be conservative, but point to areas for further research in the pursuit of greater generality for policy purposes. 

First, with regard to the data, our analysis is subject to concerns on the underlying data quality of contingent valuation studies, including hypothetical bias etc., which has been discussed at length in the literature \citep[e.g.,][]{kling2012exxon}. \cite{schlapfer2008contingent}, for instance, argues that (too) small income effects in contingent valuation studies may be an artefact of anchoring biases, but we are not aware of a clear empirical test of this hypothesis. If this were the case, we might underestimate income elasticities and, hence, the degree of complementarity, and our estimates of the appropriate upward-adjustment of natural capital values would be too conservative. Future work should thus seek to identify income elasticity also in incentivized estimation settings. 

Second, besides the specific concerns associated with contingent valuation, our approach to identifying the (aggregate) income elasticity of WTP---while building on the state of the art in the literature---is somewhat coarse, and rests on a very heterogeneous, imbalanced panel. Like \cite{heckenhahn2024relative} and contrary to other similar meta-analysis, like the one from \cite{jacobsen2009there}, we do not impose restrictions on the type of valued ecosystem service. This comprehensive approach has clear advantages, such as a very large sample size that far exceeds those of previous meta-analyses in the field. On the other hand, it raises concerns about the comparability of individual WTP estimates across a wide range of ecosystem services. This limited `commodity consistency' \citep[e.g.,][]{bergstrom2006using, moeltner2014cross} is a key feature of our analysis, as we seek to obtain an aggregate measure of the elasticity across all ecosystem services as well as measures for key ecosystem service sub-types than can be used as reasonable proxies for performing relative price change adjustments to future WTP estimates. To mitigate concerns about limited comparability---to the extent possible---we control for a number of covariates at the WTP observation-level and also present disaggregated results by the ecosystem service categories for regulating and cultural services of the \cite{assessment2005millennium} in our main results section. Still, significant heterogeneity persists due to the varied units of measurement within ecosystem service categories. Moreover, our sample contains studies that reflect both methodological refinements that have been introduced over time, which may have arguably reduced inflated WTP estimates \citep{barrio2010meta}, and an increasing share of studies from Asia and lower-income countries over time. Ideally, we'd like to identify the income elasticity of WTP based on a sample that is not subject to methodological revisions or major changes in its geographical composition. While a few test-retest investigations exist that repeatedly draw from the same sample \citep[see][for an overview]{skourtos2010reviewing}, these typically concern shorter time frames and have not been designed to investigate income effects. Evidence to date suggests that mean WTP estimates are relatively constant for up to five years, but that this is not the case for longer time frames \citep{skourtos2010reviewing}. In our meta-analysis, we find that 
the income elasticity of WTP appears relatively stable across decades. Relatedly, the underlying CV studies often elicit income only on coarse interval scales, which necessitates some subjective choice on generating an approrpiate mean income estimate. To explore sensitivity with respoct to top and bottom income observations, as well as to outliers more generally, we run further sensitivity analyses that are reported in Figures~\ref{fig:drop-top}~to~\ref{fig:drop-random} in the Appendix and show that our main estimate is reasonably robust to these variations. 

Third, our approach of relying on a direct relationship between the income elasticity of WTP and the elasticity of substitution or complementarity holds under a very common but still very specific assumption on preferences, specifically that preferences are represented by a CES utility function \citep[e.g.,][]{ebert2003environmental, baumgartner2017income}. We are not aware of studies trying to systematically test the relative goodness-of-fit of CES versus other utility specifications,\footnote{Some applied literature has documented non-constant income elasticities of WTP \citep[e.g.,][]{barbier2017income}, but no systematic evidence to date suggest a clear direction of non-constancy.} but note that extensions exist in the applied theoretical literature. One interesting case is an extension of preferences that consider critical thresholds in the form of subsistence needs \citep{baumgartner2017subsistence, drupp2018limits, heal2009climate}. If there exists some critical level of ecosystem services, $\overline E>0$, then the degree of substitutability becomes endogenous to the level of the ecosystem service over and above the critical level, and the RPC equation is adjusted to \citep[cf.][]{drupp2018limits}:\footnote{WTP estimates are typically assumed to be a function of the ecosystem service level themselves \citep{baumgartner2017income}. Empirical evidence, however, is mixed---\cite{barrio2010meta} and others find, for instance, that WTPs decrease with forest cover, while \cite{taye2021economic} find that WTPs increase with forest cover---as it's often challenging to isolate the pure effect of the level of the ecosystem service.} 
$	\quad \quad \quad RPC_t \,\,=\,\,  \xi\, \left[g_{C_t} -  g_{E_t} \, \,  \frac{E_t}{E_t -\overline E}  \right]$. 
Such an extension implies higher relative price changes that increase substantially as one gets close to the critical threshold given exogenous growth rates \citep{drupp2018limits}. It would lead to an upward revision of the natural capital values adjustment discussed in Section~\ref{sec_CWON}. However, if growth rates are endogenous and optimally managed, ensuring that we will not get close to such critical subsidence levels, relative price changes are not substantially affected \citep{drupp2021relative}.
	
Finally, we assume that preferences elicited primarily on small-scale projects aimed at improving ecosystem service conditions scale up to the global level. However, services may be perceived as complements (substitutes) at the local level, but as substitutes (complements) at a global scale. This issue may be more pronounced when the focus is relatively more on local public goods as compared to global public goods. 

We have further updated and extended the \textit{``Herculean task''} \citep[][p.\ 278]{baumgartner2015ramsey} of assembling a proxy for the aggregate growth rates of ecosystem services. There exists no accepted standard for how to aggregate various measures of environmental quality, and also the data sources we draw on have to be considered imperfect proxies themselves. We have followed~\cite{baumgartner2015ramsey} in using the unweighted arithmetic mean of the 
growth rates for the different types of ecosystem services. This assumes that the elasticity of substitution between different ecosystem services is equal to one (Cobb-Douglas), which implies that WTPs would be the same for
all types of ecosystem services if their quantities were similar, an assumption we cannot properly test. We note that there are other conceivable means of aggregation, using different weightings to different degrees of substitutability. We leave a systematic exploration of this issue to future work; the same holds for exploring the role of uncertainty around projecting past growth estimates into the future \citep{gollier2010ecological}.

Finally, our analysis has followed prior literature in estimating historical growth rates of GDP and proxies for ecosystem services while assuming that these (constant exponential) growth rates will continue into the future \citep{baumgartner2015ramsey, drupp2018limits}. In general, however, the growth rates of both goods may be interdependent and they will be endogenous to environmental management and public policy more broadly. Examples of relative price changes in the presence of an endogenous management are studied in integrated climate-economy assessment models \citep[e.g.,][]{sterner2008even, drupp2021relative, bastien2021use}. Examples of interdependencies include capitalization effects of ecosystem service changes in human-made capital and GDP. For instance, air pollution has been shown to not only harm human health but also economic productivity at a national scale \citep[e.g.,][]{fu2021air, dechezlepretre2019economic}. Reductions (improvements) in air quality may thus lead to lower (higher) GDP growth. Similarly, interdependencies arise in the case of climate regulation and its capitalization in housing prices, such as due to sea level rise, or in the case of groundwater and its capitalization in farmlands \citep{fenichel2016measuring}. If strong enough, such interdependencies can even lead to a  convergence of ecosystem service and human-made goods growth rates, as the scarcity and limited substitutability of ecosystem services as intermediate inputs to production may manifest itself as a sizable drag on growth \citep{zhu2019discounting}. This would imply that relative price changes would become smaller over time, as the limited substitutability in production would become the dominant effect. In our application (see Figure~\ref{fig:main-result-CWON}) this could be captured, in terms of its first-order effect, by considering a lower growth rate of GDP. Beyond, it would require explicit integrated modeling of the interdependencies, which is beyond the scope of our paper. 
Such as integrated analysis will have to consider that not only do ecosystem services affect economic growth, $g_C(g_E)$, as considered in \citep{zhu2019discounting}, but also that 
economic growth affects ecosystem services, $g_E(g_C)$, an effect illustrated in \citep{gollier2010ecological}. Such interdependendies may thus lead to higher or lower relative price changes as compared to the independence case illustrated in this paper.

\section{Conclusion}

We present the largest global database to estimate the degree of complementarity of ecosystem services vis-a-vis human-made goods, via the income elasticity of WTP for ecosystem services, in order to compute relative price changes of ecosystem services. We estimate an income elasticity of WTP of around \num[round-mode=places,round-precision=1]{\MainEstimate}, though this differs across ecosystem service subtypes, time frames and continents. The 95 confidence interval excludes the Cobb-Douglas case and suggests a mildly substitutive relationship between ecosystem services and market goods. This finding aligns well with the results of most similar meta-analyses, which mostly derived income elasticity estimates below unity \citep[e.g.,][]{Hokby2003, Liu2008, jacobsen2009there, Lindhjem2012, subroy2019worth}.  However, it contrasts with complementarity assumptions made in applied modelling \citep [e.g.,] []{sterner2008even}. 

For our aggregate assessment of ecosystem services, including estimates of growth rates, we find relative price changes of ecosystem services of around \num[round-mode=places,round-precision=1]{\RPCAggregateServices} percent per year. Relative price changes are smaller (\num[round-mode=places,round-precision=1]{\RPCForestServices} percent) for forest ecosystem services as these show a slower rate of de-growth as compared to other ecosystem service components. We also developed a simple approach for how these estimates can be employed to adjust future WTP estimates and present values to be used in environmental-economic accounting as demonstrated here, or in project appraisal \citep[as subsequently used in][]{drupp2024accounting}. In an application on natural capital valuation, taking the \textit{Changing Wealth of Nations (CWON)} 2021 report by the \cite{world2021changing} as a case study, we show that adjusting natural capital estimates for non-timber ecosystem services for relative price changes results in uplifting the present value over a 100 year time period by around 50 percent, materially elevating the role of public natural capital. The corresponding estimates for relative price adjustments for our aggregate assessment of public natural capital are more substantial, amounting to between about 43 and 71 percent for our main estimate of the income elasticity, depending on the social discount rate used. This echoes work on the importance of limited substitutability in climate policy appraisal \citep{bastien2021use, drupp2021relative, sterner2008even}.

While the adjustment techniques we present are generally applicable for environmental-economic accounting as well as for project appraisal, the specific numerical inputs, such as elasticites or growth rates, need to be adjusted on a case-by-case basis. We have shown, among others, that elasticities show non-negligible variation across ecosystem service types,  across continents and cross-country income levels. We hope that this spurs further investigations to assess the heterogeneity in income and substitution elasticities, as well as of good-specific growth rates, more broadly. 

Overall, our results suggest that the case for making relative price change adjustments is reasonably robust and that more countries and institutions than present \citep[][]{groom2022future} should consider making such adjustments to correct the current mis-valuation of non-market goods in public policy appraisal and of public natural capital values in comprehensive wealth accounting.

\vspace{0.5cm}
\newpage

\bibliographystyle{chicago}
\bibliography{References}

\begin{thebibliography}{}

\bibitem[\protect\citeauthoryear{Adams, Soto, Lai, Escobedo, Alvarez, and Kibria}{Adams et~al.}{2020}]{Adams2020}
Adams, D., J.~Soto, J.~Lai, F.~Escobedo, S.~Alvarez, and A.~Kibria (2020).
\newblock Public preferences and willingness to pay for invasive forest pest prevention programs in urban areas.
\newblock {\em Forests\/}~{\em 11\/}(10), 1--16.

\bibitem[\protect\citeauthoryear{Al-Assaf}{Al-Assaf}{2015}]{alassaf2015applying}
Al-Assaf, A.~A. (2015).
\newblock Applying contingent valuation to measure the economic value of forest services: A case study in northern jordan.
\newblock {\em International Journal of Sustainable Development and World Ecology\/}~{\em 22\/}(3), 242--250.

\bibitem[\protect\citeauthoryear{Barbier, Czajkowski, and Hanley}{Barbier et~al.}{2017}]{barbier2017income}
Barbier, E.~B., M.~Czajkowski, and N.~Hanley (2017).
\newblock Is the income elasticity of the willingness to pay for pollution control constant?
\newblock {\em Environmental and Resource Economics\/}~{\em 68}, 663--682.

\bibitem[\protect\citeauthoryear{Barrio and Loureiro}{Barrio and Loureiro}{2010}]{barrio2010meta}
Barrio, M. and M.~L. Loureiro (2010).
\newblock A meta-analysis of contingent valuation forest studies.
\newblock {\em Ecological Economics\/}~{\em 69\/}(5), 1023--1030.

\bibitem[\protect\citeauthoryear{Bastien-Olvera and Moore}{Bastien-Olvera and Moore}{2021}]{bastien2021use}
Bastien-Olvera, B.~A. and F.~C. Moore (2021).
\newblock Use and non-use value of nature and the social cost of carbon.
\newblock {\em Nature Sustainability\/}~{\em 4\/}(2), 101--108.

\bibitem[\protect\citeauthoryear{Baumg{\"a}rtner, Drupp, Meya, Munz, and Quaas}{Baumg{\"a}rtner et~al.}{2017}]{baumgartner2017income}
Baumg{\"a}rtner, S., M.~A. Drupp, J.~N. Meya, J.~M. Munz, and M.~F. Quaas (2017).
\newblock Income inequality and willingness to pay for environmental public goods.
\newblock {\em Journal of Environmental Economics and Management\/}~{\em 85}, 35--61.

\bibitem[\protect\citeauthoryear{Baumg{\"a}rtner, Drupp, and Quaas}{Baumg{\"a}rtner et~al.}{2017}]{baumgartner2017subsistence}
Baumg{\"a}rtner, S., M.~A. Drupp, and M.~F. Quaas (2017).
\newblock Subsistence, substitutability and sustainability in consumption.
\newblock {\em Environmental and Resource Economics\/}~{\em 67\/}(1), 47--66.

\bibitem[\protect\citeauthoryear{Baumg{\"a}rtner, Klein, Thiel, and Winkler}{Baumg{\"a}rtner et~al.}{2015}]{baumgartner2015ramsey}
Baumg{\"a}rtner, S., A.~M. Klein, D.~Thiel, and K.~Winkler (2015).
\newblock Ramsey discounting of ecosystem services.
\newblock {\em Environmental and Resource Economics\/}~{\em 61\/}(2), 273--296.

\bibitem[\protect\citeauthoryear{Bergstrom and Taylor}{Bergstrom and Taylor}{2006}]{bergstrom2006using}
Bergstrom, J.~C. and L.~O. Taylor (2006).
\newblock Using meta-analysis for benefits transfer: Theory and practice.
\newblock {\em Ecological economics\/}~{\em 60\/}(2), 351--360.

\bibitem[\protect\citeauthoryear{Bliem and Getzner}{Bliem and Getzner}{2012}]{bliem2012willingness}
Bliem, M. and M.~Getzner (2012).
\newblock Willingness-to-pay for river restoration: Differences across time and scenarios.
\newblock {\em Environmental Economics and Policy Studies\/}~{\em 14\/}(3), 241--260.

\bibitem[\protect\citeauthoryear{Butchart, Walpole, Collen, Van~Strien, Scharlemann, Almond, Baillie, Bomhard, Brown, Bruno, et~al.}{Butchart et~al.}{2010}]{butchart2010global}
Butchart, S.~H., M.~Walpole, B.~Collen, A.~Van~Strien, J.~P. Scharlemann, R.~E. Almond, J.~E. Baillie, B.~Bomhard, C.~Brown, J.~Bruno, et~al. (2010).
\newblock Global biodiversity: indicators of recent declines.
\newblock {\em Science\/}~{\em 328\/}(5982), 1164--1168.

\bibitem[\protect\citeauthoryear{Cohen, Hepburn, and Teytelboym}{Cohen et~al.}{2019}]{cohen2019natural}
Cohen, F., C.~J. Hepburn, and A.~Teytelboym (2019).
\newblock Is natural capital really substitutable?
\newblock {\em Annual Review of Environment and Resources\/}~{\em 44}, 425--448.

\bibitem[\protect\citeauthoryear{Dasgupta}{Dasgupta}{2021}]{dasgupta2021economics}
Dasgupta, P. (2021).
\newblock The economics of biodiversity and ecosystem services: The dasgupta review.

\bibitem[\protect\citeauthoryear{Dechezlepr{\^e}tre, Rivers, and Stadler}{Dechezlepr{\^e}tre et~al.}{2019}]{dechezlepretre2019economic}
Dechezlepr{\^e}tre, A., N.~Rivers, and B.~Stadler (2019).
\newblock The economic cost of air pollution: Evidence from europe.

\bibitem[\protect\citeauthoryear{Dogan and Muhammad}{Dogan and Muhammad}{2019}]{dogan2019willingness}
Dogan, E. and I.~Muhammad (2019).
\newblock Willingness to pay for renewable electricity: A contingent valuation study in turkey.
\newblock {\em Electricity Journal\/}~{\em 32\/}(10).

\bibitem[\protect\citeauthoryear{Drupp}{Drupp}{2018}]{drupp2018limits}
Drupp, M.~A. (2018).
\newblock Limits to substitution between ecosystem services and manufactured goods and implications for social discounting.
\newblock {\em Environmental and Resource Economics\/}~{\em 69\/}(1), 135--158.

\bibitem[\protect\citeauthoryear{Drupp, Baumg{\"a}rtner, Meyer, Quaas, and von Wehrden}{Drupp et~al.}{2020}]{drupp2020between}
Drupp, M.~A., S.~Baumg{\"a}rtner, M.~Meyer, M.~F. Quaas, and H.~von Wehrden (2020).
\newblock Between ostrom and nordhaus: The research landscape of sustainability economics.
\newblock {\em Ecological Economics\/}~{\em 172}, 106620.

\bibitem[\protect\citeauthoryear{Drupp, Freeman, Groom, and Nesje}{Drupp et~al.}{2018}]{drupp2018discounting}
Drupp, M.~A., M.~C. Freeman, B.~Groom, and F.~Nesje (2018).
\newblock Discounting disentangled.
\newblock {\em American Economic Journal: Economic Policy\/}~{\em 10\/}(4), 109--34.

\bibitem[\protect\citeauthoryear{Drupp and H{\"a}nsel}{Drupp and H{\"a}nsel}{2021}]{drupp2021relative}
Drupp, M.~A. and M.~C. H{\"a}nsel (2021).
\newblock Relative prices and climate policy: How the scarcity of nonmarket goods drives policy evaluation.
\newblock {\em American Economic Journal: Economic Policy\/}~{\em 13\/}(1), 168--201.

\bibitem[\protect\citeauthoryear{Drupp, H{\"a}nsel, Fenichel, Freeman, Gollier, Groom, Heal, Howard, Millner, Moore, Nesje, Quaas, Smulders, Sterner, Traeger, and Venmans}{Drupp et~al.}{2024}]{drupp2024accounting}
Drupp, M.~A., M.~C. H{\"a}nsel, E.~P. Fenichel, M.~Freeman, C.~Gollier, B.~Groom, G.~M. Heal, P.~H. Howard, A.~Millner, F.~C. Moore, F.~Nesje, M.~F. Quaas, S.~Smulders, T.~Sterner, C.~Traeger, and F.~Venmans (2024).
\newblock Accounting for the increasing benefits from scarce ecosystems.
\newblock {\em Science\/}~{\em 383\/}(6687), 1062--1064.

\bibitem[\protect\citeauthoryear{Ebert}{Ebert}{2003}]{ebert2003environmental}
Ebert, U. (2003).
\newblock Environmental goods and the distribution of income.
\newblock {\em Environmental and Resource Economics\/}~{\em 25\/}(4), 435--459.

\bibitem[\protect\citeauthoryear{Endalew, Wondimagegnhu, and Tassie}{Endalew et~al.}{2020}]{endalew2020willingness}
Endalew, B., B.~A. Wondimagegnhu, and K.~Tassie (2020).
\newblock Willingness to pay for church forest conservation: A case study in northwestern ethiopia.
\newblock {\em Journal of Forest Science\/}~{\em 66\/}(3), 105--116.

\bibitem[\protect\citeauthoryear{Fenichel, Abbott, Bayham, Boone, Haacker, and Pfeiffer}{Fenichel et~al.}{2016}]{fenichel2016measuring}
Fenichel, E.~P., J.~K. Abbott, J.~Bayham, W.~Boone, E.~M. Haacker, and L.~Pfeiffer (2016).
\newblock Measuring the value of groundwater and other forms of natural capital.
\newblock {\em Proceedings of the National Academy of Sciences\/}~{\em 113\/}(9), 2382--2387.

\bibitem[\protect\citeauthoryear{Fu, Viard, and Zhang}{Fu et~al.}{2021}]{fu2021air}
Fu, S., V.~B. Viard, and P.~Zhang (2021).
\newblock Air pollution and manufacturing firm productivity: Nationwide estimates for china.
\newblock {\em The Economic Journal\/}~{\em 131\/}(640), 3241--3273.

\bibitem[\protect\citeauthoryear{Gollier}{Gollier}{2010}]{gollier2010ecological}
Gollier, C. (2010).
\newblock Ecological discounting.
\newblock {\em Journal of economic theory\/}~{\em 145\/}(2), 812--829.

\bibitem[\protect\citeauthoryear{Groom, Drupp, Freeman, and Nesje}{Groom et~al.}{2022}]{groom2022future}
Groom, B., M.~Drupp, M.~C. Freeman, and F.~Nesje (2022).
\newblock The future, now: A review of social discounting.
\newblock {\em Annual Review of Resource Economics\/}.

\bibitem[\protect\citeauthoryear{Groom and Hepburn}{Groom and Hepburn}{2017}]{groom2017reflections}
Groom, B. and C.~Hepburn (2017).
\newblock Reflections—looking back at social discounting policy: the influence of papers, presentations, political preconditions, and personalities.
\newblock {\em Review of Environmental Economics and Policy\/}~{\em 11\/}(2), 336--356.

\bibitem[\protect\citeauthoryear{Guo, Wang, and Zhang}{Guo et~al.}{2020}]{guo2020pollution}
Guo, D., A.~Wang, and A.~T. Zhang (2020).
\newblock Pollution exposure and willingness to pay for clean air in urban china.
\newblock {\em Journal of Environmental Management\/}~{\em 261}.

\bibitem[\protect\citeauthoryear{Hanley, Dupuy, and McLaughlin}{Hanley et~al.}{2015}]{hanleyetal2015}
Hanley, N., L.~Dupuy, and E.~McLaughlin (2015).
\newblock Genuine savings and sustainability.
\newblock {\em Journal of Economic Surveys\/}~{\em 29\/}(4), 779--806.

\bibitem[\protect\citeauthoryear{Heal}{Heal}{2009}]{heal2009climate}
Heal, G. (2009).
\newblock Climate economics: a meta-review and some suggestions for future research.
\newblock {\em Review of Environmental economics and Policy\/}.

\bibitem[\protect\citeauthoryear{Heckenhahn and Drupp}{Heckenhahn and Drupp}{2024}]{heckenhahn2024relative}
Heckenhahn, J. and M.~A. Drupp (2024).
\newblock Relative price changes of ecosystem services: Evidence from germany.
\newblock {\em Environmental and Resource Economics\/}~{\em 87}, 833–--880.

\bibitem[\protect\citeauthoryear{Hoel and Sterner}{Hoel and Sterner}{2007}]{hoel2007discounting}
Hoel, M. and T.~Sterner (2007).
\newblock Discounting and relative prices.
\newblock {\em Climatic Change\/}~{\em 84\/}(3), 265--280.

\bibitem[\protect\citeauthoryear{Huenchuleo, Barkmann, and Villalobos}{Huenchuleo et~al.}{2012}]{huenchuleo2012social_psychology}
Huenchuleo, C., J.~Barkmann, and P.~Villalobos (2012).
\newblock Social psychology predictors for the adoption of soil conservation measures in central chile.
\newblock {\em Land Degradation and Development\/}~{\em 23\/}(5), 483--495.

\bibitem[\protect\citeauthoryear{Hökby and Söderqvist}{Hökby and Söderqvist}{2003}]{Hokby2003}
Hökby, S. and T.~Söderqvist (2003).
\newblock Elasticities of demand and willingness to pay for environmental services in sweden.
\newblock {\em Environmental and Resource Economics\/}~{\em 26\/}(3), 361--383.

\bibitem[\protect\citeauthoryear{IPBES}{IPBES}{2019}]{IPBES2019}
IPBES (2019).
\newblock Summary for policymakers of the global assessment report on biodiversity and ecosystem services of the intergovernmental science-policy platform on biodiversity and ecosystem services.
\newblock Technical report, Intergovernmental Platform on Biodiversity and Ecosystem Services.

\bibitem[\protect\citeauthoryear{Jacobsen and Hanley}{Jacobsen and Hanley}{2009}]{jacobsen2009there}
Jacobsen, J.~B. and N.~Hanley (2009).
\newblock Are there income effects on global willingness to pay for biodiversity conservation?
\newblock {\em Environmental and Resource Economics\/}~{\em 43\/}(2), 137--160.

\bibitem[\protect\citeauthoryear{Kling, Phaneuf, and Zhao}{Kling et~al.}{2012}]{kling2012exxon}
Kling, C.~L., D.~J. Phaneuf, and J.~Zhao (2012).
\newblock From exxon to bp: Has some number become better than no number?
\newblock {\em Journal of Economic Perspectives\/}~{\em 26\/}(4), 3--26.

\bibitem[\protect\citeauthoryear{Lindhjem and Tuan}{Lindhjem and Tuan}{2012}]{Lindhjem2012}
Lindhjem, H. and T.~L. Tuan (2012).
\newblock Valuation of ecosystem services in a green economy: Willingness to pay for forest conservation in vietnam.
\newblock {\em Environmental Economics and Policy Studies\/}~{\em 14\/}(4), 303--323.

\bibitem[\protect\citeauthoryear{Liu and Stern}{Liu and Stern}{2008}]{Liu2008}
Liu, S. and D.~I. Stern (2008).
\newblock A meta-analysis of contingent valuation studies in coastal and near-shore marine ecosystems.
\newblock {\em Marine Resource Economics\/}~{\em 23\/}(2), 119--147.

\bibitem[\protect\citeauthoryear{Ma, Min, Xu, and Sang}{Ma et~al.}{2021}]{ma2021ecotourism_china}
Ma, T., Q.~Min, K.~Xu, and W.~Sang (2021).
\newblock Resident willingness to pay for ecotourism resources and associated factors in sanjiangyuan national park, china.
\newblock {\em Journal of Resources and Ecology\/}~{\em 12\/}(5), 693--706.

\bibitem[\protect\citeauthoryear{Maharana, Rai, and Sharma}{Maharana et~al.}{2000}]{maharana2000valuing_ecotourism_sikkim}
Maharana, I., S.~Rai, and E.~Sharma (2000).
\newblock Valuing ecotourism in a sacred lake of the sikkim himalaya, india.
\newblock {\em Environmental Conservation\/}~{\em 27\/}(3), 269--277.

\bibitem[\protect\citeauthoryear{MEA}{MEA}{2005}]{assessment2005millennium}
MEA (2005).
\newblock {\em Millennium ecosystem assessment}.
\newblock Millennium Ecosystem Assessment.

\bibitem[\protect\citeauthoryear{Meyerhoff, Angeli, and Hartje}{Meyerhoff et~al.}{2012}]{meyerhoff2012valuing_benefits}
Meyerhoff, J., D.~Angeli, and V.~Hartje (2012).
\newblock Valuing the benefits of implementing a national strategy on biological diversity—the case of germany.
\newblock {\em Environmental Science and Policy\/}~{\em 23}, 109--119.

\bibitem[\protect\citeauthoryear{Moeltner and Rosenberger}{Moeltner and Rosenberger}{2014}]{moeltner2014cross}
Moeltner, K. and R.~S. Rosenberger (2014).
\newblock Cross-context benefit transfer: A bayesian search for information pools.
\newblock {\em American Journal of Agricultural Economics\/}~{\em 96\/}(2), 469--488.

\bibitem[\protect\citeauthoryear{Moore, Drupp, Rising, Dietz, Rudik, and Wagner}{Moore et~al.}{2024}]{moore2024synthesis}
Moore, F.~C., M.~A. Drupp, J.~Rising, S.~Dietz, I.~Rudik, and G.~Wagner (2024).
\newblock Synthesis of evidence yields high social cost of carbon due to structural model variation and uncertainties.
\newblock {\em NBER Working Paper\/}~{\em 32544}.

\bibitem[\protect\citeauthoryear{Mwebaze, Marris, Brown, MacLeod, Jones, and Budge}{Mwebaze et~al.}{2018}]{mwebaze2018bee_pollination}
Mwebaze, P., G.~Marris, M.~Brown, A.~MacLeod, G.~Jones, and G.~Budge (2018).
\newblock Measuring public perception and preferences for ecosystem services: A case study of bee pollination in the uk.
\newblock {\em Land Use Policy\/}~{\em 71}, 355--362.

\bibitem[\protect\citeauthoryear{Neumayer}{Neumayer}{2003}]{neumayer2003weak}
Neumayer, E. (2003).
\newblock {\em Weak versus strong sustainability: exploring the limits of two opposing paradigms}.
\newblock Edward Elgar Publishing.

\bibitem[\protect\citeauthoryear{Petrolia and Kim}{Petrolia and Kim}{2009}]{petrolia2009barrier_islands}
Petrolia, D. and T.-G. Kim (2009).
\newblock What are barrier islands worth? estimates of willingness to pay for restoration.
\newblock {\em Marine Resource Economics\/}~{\em 24\/}(2), 131--146.

\bibitem[\protect\citeauthoryear{Ready, Malzubris, and Senkane}{Ready et~al.}{2002}]{ready2002relationship}
Ready, R.~C., J.~Malzubris, and S.~Senkane (2002).
\newblock The relationship between environmental values and income in a transition economy: surface water quality in latvia.
\newblock {\em Environment and Development Economics\/}~{\em 7\/}(1), 147--156.

\bibitem[\protect\citeauthoryear{Richardson and Loomis}{Richardson and Loomis}{2009}]{richardson2009total}
Richardson, L. and J.~Loomis (2009).
\newblock The total economic value of threatened, endangered and rare species: an updated meta-analysis.
\newblock {\em Ecological economics\/}~{\em 68\/}(5), 1535--1548.

\bibitem[\protect\citeauthoryear{Rouhi~Rad, Adamowicz, Entem, Fenichel, and Lloyd-Smith}{Rouhi~Rad et~al.}{2021}]{rouhi2021complementarity}
Rouhi~Rad, M., W.~Adamowicz, A.~Entem, E.~P. Fenichel, and P.~Lloyd-Smith (2021).
\newblock Complementarity (not substitution) between natural and produced capital: Evidence from the panama canal expansion.
\newblock {\em Journal of the Association of Environmental and Resource Economists\/}~{\em 8\/}(6), 1115--1146.

\bibitem[\protect\citeauthoryear{Schl{\"a}pfer}{Schl{\"a}pfer}{2008}]{schlapfer2008contingent}
Schl{\"a}pfer, F. (2008).
\newblock Contingent valuation: a new perspective.
\newblock {\em Ecological economics\/}~{\em 64\/}(4), 729--740.

\bibitem[\protect\citeauthoryear{Schläpfer and Getzner}{Schläpfer and Getzner}{2020}]{schlapfer2020behavioral_economics}
Schläpfer, F. and M.~Getzner (2020).
\newblock Beyond current guidelines: A proposal for bringing behavioral economics to the design and analysis of stated preference surveys.
\newblock {\em Ecological Economics\/}~{\em 176}.

\bibitem[\protect\citeauthoryear{Siikam{\"a}ki, Piaggio, da~Silva, {\'A}lvarez, and Chu}{Siikam{\"a}ki et~al.}{2021}]{siikamaki2021global}
Siikam{\"a}ki, J., M.~Piaggio, N.~da~Silva, I.~{\'A}lvarez, and Z.~Chu (2021).
\newblock Global assessment of non-wood forest ecosystem services: A revision of a spatially explicit meta-analysis and benefit transfer.

\bibitem[\protect\citeauthoryear{Siikam{\"a}ki, Santiago-{\'A}vila, and Vail}{Siikam{\"a}ki et~al.}{2015}]{siikamaki2015global}
Siikam{\"a}ki, J., F.~J. Santiago-{\'A}vila, and P.~Vail (2015).
\newblock Global assessment of non-wood forest ecosystem services.
\newblock {\em Spatially Explicit Meta-Analysis and Benefit Transfer to Improve the World Bank’s Forest Weatlh Database\/}, 1--97.

\bibitem[\protect\citeauthoryear{Skourtos, Kontogianni, and Harrison}{Skourtos et~al.}{2010}]{skourtos2010reviewing}
Skourtos, M., A.~Kontogianni, and P.~Harrison (2010).
\newblock Reviewing the dynamics of economic values and preferences for ecosystem goods and services.
\newblock {\em Biodiversity and conservation\/}~{\em 19\/}(10), 2855--2872.

\bibitem[\protect\citeauthoryear{Smith}{Smith}{2023}]{smith2023accounting}
Smith, V.~K. (2023).
\newblock Accounting for income inequality in benefit transfers: The importance of the income elasticity of wtp.
\newblock {\em Journal of Environmental Economics and Management\/}, 102781.

\bibitem[\protect\citeauthoryear{Smulders and van Soest}{Smulders and van Soest}{2023}]{smulders2023}
Smulders, S. and D.~van Soest (2023).
\newblock Natural capital substitution: Implications for growth, shadow prices, and natural capital accounting.
\newblock {\em Mimeo, Tilburg University\/}.

\bibitem[\protect\citeauthoryear{Sterner and Persson}{Sterner and Persson}{2008}]{sterner2008even}
Sterner, T. and U.~M. Persson (2008).
\newblock An even sterner review: Introducing relative prices into the discounting debate.
\newblock {\em Review of Environmental Economics and Policy\/}.

\bibitem[\protect\citeauthoryear{Subroy, Gunawardena, Polyakov, Pandit, and Pannell}{Subroy et~al.}{2019}]{subroy2019worth}
Subroy, V., A.~Gunawardena, M.~Polyakov, R.~Pandit, and D.~J. Pannell (2019).
\newblock The worth of wildlife: A meta-analysis of global non-market values of threatened species.
\newblock {\em Ecological Economics\/}~{\em 164}, 106374.

\bibitem[\protect\citeauthoryear{Taye, Folkersen, Fleming, Buckwell, Mackey, Diwakar, Le, Hasan, and Saint~Ange}{Taye et~al.}{2021}]{taye2021economic}
Taye, F.~A., M.~V. Folkersen, C.~M. Fleming, A.~Buckwell, B.~Mackey, K.~Diwakar, D.~Le, S.~Hasan, and C.~Saint~Ange (2021).
\newblock The economic values of global forest ecosystem services: A meta-analysis.
\newblock {\em Ecological Economics\/}~{\em 189}, 107145.

\bibitem[\protect\citeauthoryear{Tolunay and Başsüllü}{Tolunay and Başsüllü}{2015}]{tolunay2015wtp_carbon_sequestration}
Tolunay, A. and C.~Başsüllü (2015).
\newblock Willingness to pay for carbon sequestration and co-benefits of forests in turkey.
\newblock {\em Sustainability\/}~{\em 7\/}(3), 3311--3337.

\bibitem[\protect\citeauthoryear{Traeger}{Traeger}{2011}]{traeger2011sustainability}
Traeger, C.~P. (2011).
\newblock Sustainability, limited substitutability, and non-constant social discount rates.
\newblock {\em Journal of Environmental Economics and Management\/}~{\em 62\/}(2), 215--228.

\bibitem[\protect\citeauthoryear{Treasury}{Treasury}{2021}]{HMT2021_Environmental}
Treasury, H. (2021).
\newblock Green book supplementary document: Environmental discount rate review, conclusion.
\newblock {\em HM Treasury\/}.

\bibitem[\protect\citeauthoryear{Tziakis, Pachiadakis, Moraitakis, Xideas, Theologis, and Tsagarakis}{Tziakis et~al.}{2009}]{tziakis2009wtp_wastewater_treatment}
Tziakis, I., I.~Pachiadakis, M.~Moraitakis, K.~Xideas, G.~Theologis, and K.~Tsagarakis (2009).
\newblock Valuing benefits from wastewater treatment and reuse using contingent valuation methodology.
\newblock {\em Desalination\/}~{\em 237}, --.

\bibitem[\protect\citeauthoryear{Weikard and Zhu}{Weikard and Zhu}{2005}]{weikard2005discounting}
Weikard, H.-P. and X.~Zhu (2005).
\newblock Discounting and environmental quality: When should dual rates be used?
\newblock {\em Economic Modelling\/}~{\em 22\/}(5), 868--878.

\bibitem[\protect\citeauthoryear{World~Bank}{World~Bank}{2021}]{world2021changing}
World~Bank, T. (2021).
\newblock {\em The Changing Wealth of Nations 2021: Managing Assets for the Future}.
\newblock The World Bank.

\bibitem[\protect\citeauthoryear{Yang, Zou, Lin, Wu, and Wang}{Yang et~al.}{2014}]{yang2014wtp_co2_mitigation}
Yang, J., L.~Zou, T.~Lin, Y.~Wu, and H.~Wang (2014).
\newblock Public willingness to pay for co2 mitigation and the determinants under climate change: A case study of suzhou, china.
\newblock {\em Journal of Environmental Management\/}~{\em 146}, 1--8.

\bibitem[\protect\citeauthoryear{Zhu, Smulders, and de~Zeeuw}{Zhu et~al.}{2019}]{zhu2019discounting}
Zhu, X., S.~Smulders, and A.~de~Zeeuw (2019).
\newblock Discounting in the presence of scarce ecosystem services.
\newblock {\em Journal of Environmental Economics and Management\/}~{\em 98}, 102--272.

\end{thebibliography}

\newpage

\appendix
\begin{appendices}

	\cleardoublepage
	\thispagestyle{empty}
	\setcounter{page}{0}

 	\noindent \textbf{\Huge{Appendix}}

\section{Selection of relevant valuation studies}

 
\subsection{Search string \label{Search string}}

Our focus is on values for regulating ecosystem services and cultural ecosystem services (not provisioning services) that have been elicited using the contingent valuation method. The search string has three components (1) focus on ecosystem services, (2) focus on WTP estimates, (3) focus on the contingent valuation method.

( TITLE-ABS-KEY ( environment*  OR natur* OR  ecosystem  OR  biodiversity  OR biologic* OR  ecologic* OR  habitat*  OR  forest* OR  species  OR  protected OR  conserv*  OR  endangered OR  "national park*"  OR  landscape*  OR  terrestrial  OR  pollination  OR  tree*  OR  tropic*  OR  vegetation  OR  peatland*  OR grassland* OR dryland* OR pastoral OR  soil  OR  animal*  OR  bird*  OR  wild*  OR  air OR  water OR  aquatic  OR  marine  OR coast* OR  water*  OR  fish*  OR  wetland*  OR  mangrove*  OR  reef*  OR  marsh*  OR  floodplain*  OR  river*  OR  climate  OR  storm*  OR  erosion OR  pest*  OR  hazard* OR  recreat*  OR  touris*  OR “urban green” OR sacred OR spirit* OR sanctuary OR “natural heritage” OR aesthetic*)  

AND  TITLE-ABS-KEY ( wtp  OR  willingness-to-pay  OR  "willingness to pay*"  OR  "willing to pay*"  OR  "shadow price*"  OR  "shadow value*"  OR  "implicit price*"  OR  "implicit value*")  

AND  TITLE-ABS-KEY ( "contingent valuation*"  OR  cvm  OR  "contingent choice*") )  

AND  ( LIMIT-TO ( SRCTYPE ,  "j" ) )  AND  ( LIMIT-TO ( DOCTYPE ,  "ar" ) )  AND  ( LIMIT-TO ( PUBYEAR ,  2021 )  OR  LIMIT-TO ( PUBYEAR ,  2020 )  OR  LIMIT-TO ( PUBYEAR ,  2019 )  OR  LIMIT-TO ( PUBYEAR ,  2018 )  OR  LIMIT-TO ( PUBYEAR ,  2017 )  OR  LIMIT-TO ( PUBYEAR ,  2016 )  OR  LIMIT-TO ( PUBYEAR ,  2015 )  OR  LIMIT-TO ( PUBYEAR ,  2014 )  OR  LIMIT-TO ( PUBYEAR ,  2013 )  OR  LIMIT-TO ( PUBYEAR ,  2012 )  OR  LIMIT-TO ( PUBYEAR ,  2011 )  OR  LIMIT-TO ( PUBYEAR ,  2010 )  OR  LIMIT-TO ( PUBYEAR ,  2009 )  OR  LIMIT-TO ( PUBYEAR ,  2008 )  OR  LIMIT-TO ( PUBYEAR ,  2007 )  OR  LIMIT-TO ( PUBYEAR ,  2006 )  OR  LIMIT-TO ( PUBYEAR ,  2005 )  OR  LIMIT-TO ( PUBYEAR ,  2004 )  OR  LIMIT-TO ( PUBYEAR ,  2003 )  OR  LIMIT-TO ( PUBYEAR ,  2002 )  OR  LIMIT-TO ( PUBYEAR ,  2001 )  OR  LIMIT-TO ( PUBYEAR ,  2000 ) )  AND  ( LIMIT-TO ( LANGUAGE ,  "English" ) ) 

\newpage
\subsection{Exclusion and selection criteria \label{Exclusion criteria}}

\subsubsection{Paper exclusion criteria}

Citations:
    We excluded all studies that had not been cited (in SCOPUS).

\noindent Abstract screening: We excluded non-topical publications based on abstract-screening that do not report new primary WTP estimates. Specifically, we excluded: Theory, reviews, comments, non-primary valuation (such as benefit transfer), as well as WTPs for non-environmental goods, WTPs for provisioning services, WTPs derived from valuation approaches other than CV 

\noindent PDFs obtainable: We excluded studies where we could not access the PDFs. 

\noindent Paper screening: We excluded non-topical publications based on abstract-screening that do not report new primary WTP estimates. Specifically, we excluded: Theory, reviews, comments, non-primary valuation (such as benefit transfer), as well as WTPs for non-environmental goods, WTPs for provisioning services, WTPs derived from valuation approaches other than CV.

Overall, our approach to WTP studies is intentionally inclusive to ensure that our meta-analysis accurately captures the full range of values associated with ecosystem services. Some included studies may not explicitly refer to ecosystem services, however, in the respective cases, we deemed it reasonable to assume that study participants associated ecosystem services with the value of the goods being assessed (e.g., we generally assumed renewable energy to be associated with climate regulation in people's minds).\footnote{The decision for inclusion was particularly challenging for studies estimating non-market ecosystem service values embedded in otherwise market goods. For example, in Kim et al. (2019) and Milovantseva (2016), non-market ecosystem service values related to climate and air quality regulation, and water pollution and waste regulation, respectively, are suggested during the creation of market goods - cell phones. The authors, however, take care to isolate ecosystem service-relevant components by framing their survey prompts relative to baselines of otherwise similar devices without the ecosystem service component. On this basis, we decided to include these studies in our analysis.}

\subsubsection{Data selection criteria}

In the following, we detail our approach for selecting WTP and income values, which constitute the key variables for our analyses.

WTP data selection: We exclude median WTP values, WTP values derived from multiplying marginal WTP estimates, WTP values resulting from the addition of preceding WTP values, WTP values based on pretests, WTP values based on subsamples when overall mean values are provided, and per-use WTP values. When different results are presented based on different models, we include only the WTP values from the standard model. If no standard model is indicated, we average the relevant model results. When multiple mean WTP estimates are provided (e.g., including or excluding outliers and zero bids), we include the estimate marked as the authors' preferred estimate. If no preference is indicated, we include the unmodified estimate. When WTP values are provided for different subsamples, we assign the WTP values to the corresponding subsample income values. When WTP values refer to a monthly payment, we multiply these values by 12 to obtain annual values. WTP values referring to yearly payments and one-time payments are included as they are. When WTP results are divided among different quantities (supply levels) of the same ecosystem service, we take the most marginal of these values, though alternatives of taking their average or including all levels as separate estimates are also considered. If WTP results consider participants' response uncertainty, we average these values. When WTP results are split among different subsamples without overall mean WTP values or subsample-specific income values, we take the average of the subsample WTP values, using weighted averages if subsample sizes are available. 

The inclusion of negative WTP values results in the inclusion of one additional estimate. In the relevant study, four estimates are provided where one is negative but statistically insignificant from zero at standard levels of evaluation. Exclusion of the negative WTP estimate and inclusion by two separate approaches results in a slight impact. Our method of inclusion is by transforming the negative estimate as ${-ln(abs({WTP}_{ij}))}$ and then proceeding with our estimation strategy as in the main text. The alternative method is to substitute ${ln(0.0001)}$ for the negative estimate and include an indicator variable equal to one for the observation.

The exclusion of per-use values is necessary when studies do not---as in every case we encounter---report on a per-respondent basis how many times respondents use the studied ecosystem service over an identified period of time such as a year. The number of uses may differ substantially from any mean usage estimate. As such, the excluded  per-use estimates (around 80 observations) cannot be placed on a comparable timescale to our included estimates. 

Income data selection: We include studies regardless of whether they provide net or gross income data, while we contacted study authors when articles did not provide specific information on that. We also included studies regardless of whether the respective income data refers to the household or personal level.  If a study only provides percentage shares of income categories instead of a mean income value, we derive the mean income value by calculating the midpoints of the income categories and multiplying them by their respective percentage shares. For the category open towards the bottom, we multiply the upper bound (the lower bound of the lowest income category) by 0.75 to find the midpoint, and for the category open towards the top, we multiply the lower bound (the higher bound of the highest income category) by 1.5. We then sum these products and divide by the sum of the percentage shares to estimate the mean income. For income values split among different subsamples, we average these values to attain overall mean income values, using weighted averages if subsample sizes are available.

Recognising that the top and bottom income category multipliers (0.75 and 1.5, respectively) are judgment calls, we test the sensitivity of our results to alternatives. We do so by iterating over lower-income category multipliers of 0 to 0.9 and over upper-income category multipliers of 1.0 to 1.9 in increments of 0.1. In all cases, income elasticity of willingness to pay estimates change by less than three-percent. As the lower- and upper-income categories apply to small groups of the population in most studies---and is apply equally to all relevant studies---it is unsurprising that results are insensitive to this adjustment.


\newpage

\section{Graphical presentation of the meta-analysis data}
\label{AppendixB}

\begin{figure}[H]
    \centering
    \caption{Visualization of mean income and WTP data (original and ln-transformed)}
    \includegraphics[width=0.95\linewidth]{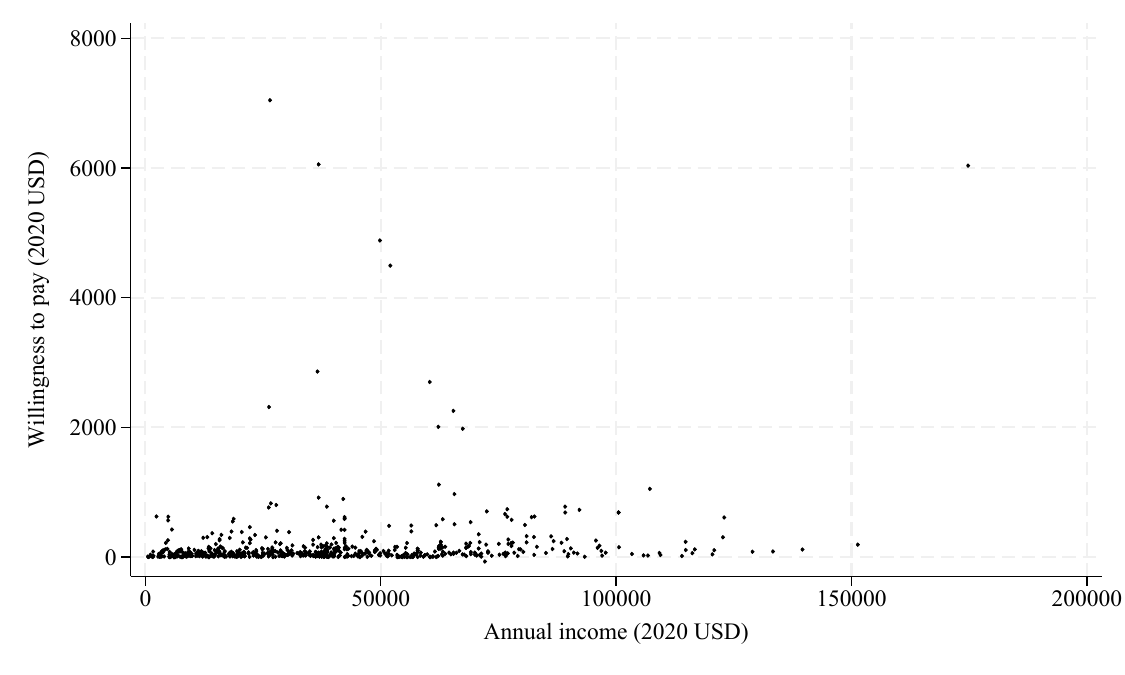}
    \label{fig:original}
        \includegraphics[width=0.95\linewidth]{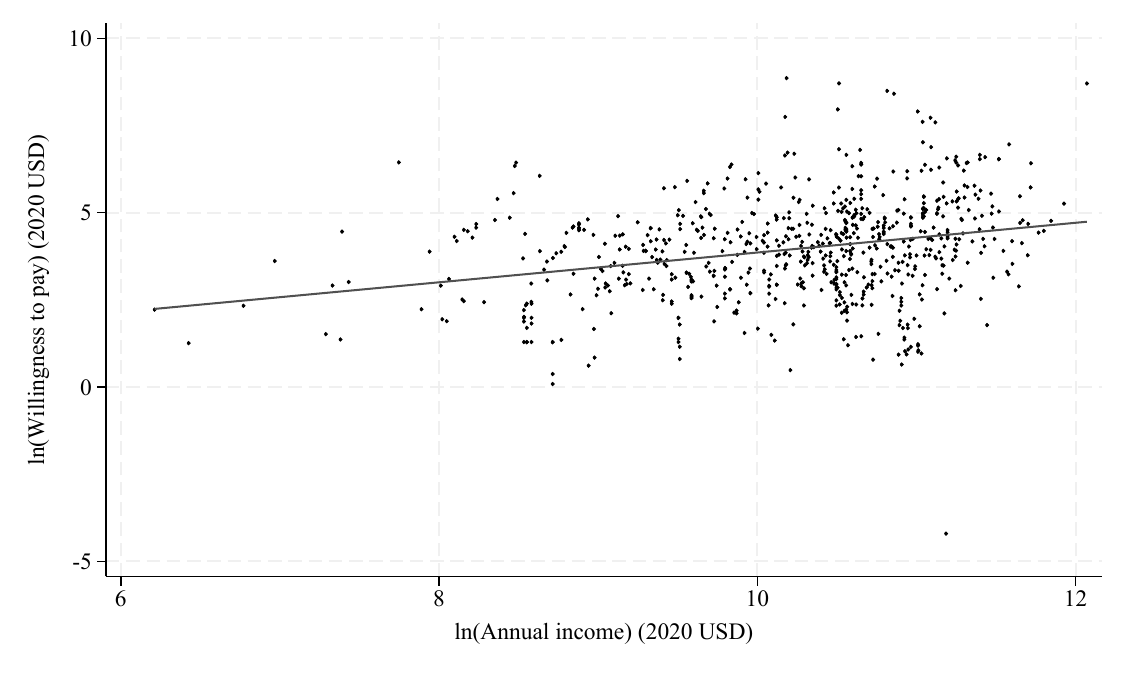}
    \label{fig:ln_transformed}
\end{figure}

Figure~\ref{fig:original} visualizes the meta-analysis data using the original, untransformed income and WTP data in the upper panel. Here, each dot represents a WTP value. In contrast, the lower panel presents both WTP and income data in their logarithmic forms, which we consistently use throughout our main analysis to calculate income elasticities. Here, each dot represents a ln(WTP) value. The lower panel also includes a regression line based on the univariate version of our preferred square root of sample size weighting regression model.

\FloatBarrier

\FloatBarrier

\section{Inflation and currency conversion}

All monetary values were converted to 2020 US Dollar by first inflating the respective national consumer price index and then applying purchasing-power-parity (PPP) conversion. The relevant year for the inflation of the values was the year of study data collection. When the authors did not provide the study year, we estimate the average lag between study and publication years based on the studies where both pieces of information is available. The difference is approximately 4.0 years on average. We use this to estimate the study year when missing. When historical inflation data for years far in the past were unavailable, we utilized the most recent year's inflation data as an estimate for these years' inflation rates.


\section{Alternative specification results \label{spec chart}}

This section presents alternative specifications to explore the robustness of our results. We first present a specification graph to suggest the robustness of our results to the inclusion or exclusion of covariates. Second, we present results based on alternative statistical models (fixed-effects, random-effects, weighted and unweighted OLS) to suggest the robust of our results to model selection.Third, we present the sensitivity of results to alternative weighting methods. Fourth, we test the sensitivity of results to dropping successively larger portions of the dataset in terms of top and bottom incomes, in turn. Finally, we test the sensitivity of results to randomly dropping successively larger shares of the dataset.

We test $2^{15}=32,768$ alternative specifications based on including or excluding variables and report the results as a specification graph. Alternatives potentially include the covariates in our main specification, plus respondent age, household size, survey format, continent, time period (pre- versus post-2011), and whether the study pertains to forests. Inclusion of the MEA list of regulating and cultural services indicators variables is treated as one group (either included or excluded together) rather than individually to avoid running alternative specification on $2^{26}$ combinations.



 \begin{figure}[h!]
   	\vspace{0.32cm}
	\caption{Income elasticity of WTP estimates based on alternative model specifications.}
	\label{fig:spec-chart_Fig5}
	\centering
	\begin{minipage}{1\textwidth} 
	    \includegraphics[width=1\linewidth]{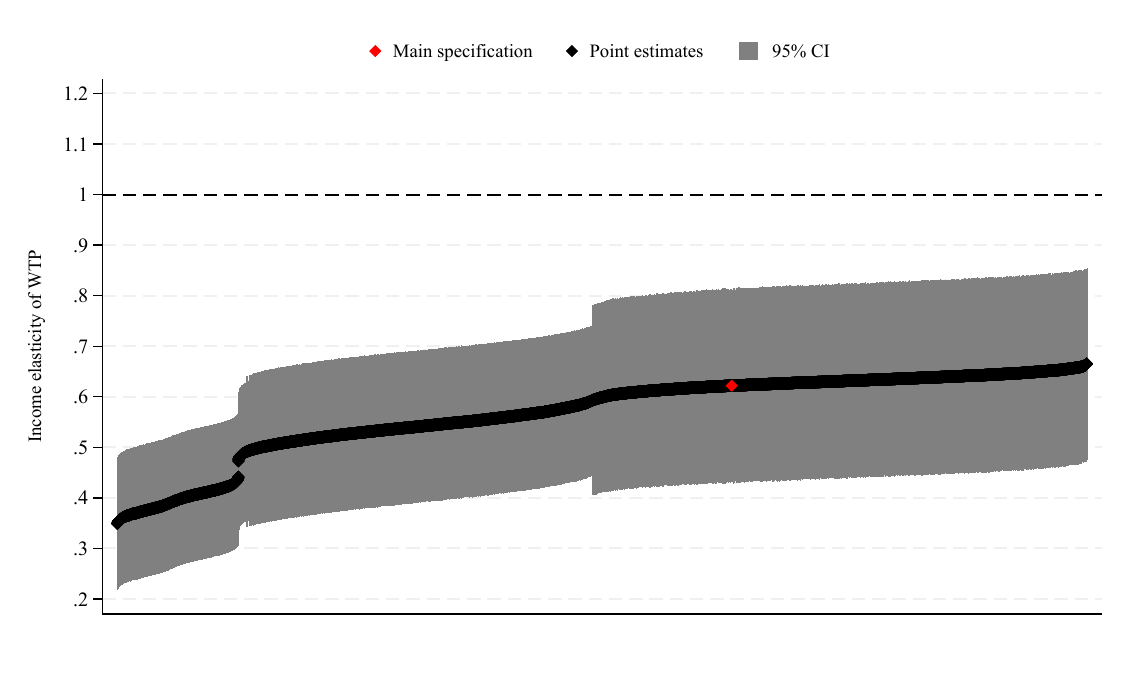 }
	\end{minipage}
 	\begin{minipage}{0.97\textwidth} 
   \flushleft{\footnotesize \emph{Notes:} Estimates are the result of $2^{15}=32,768$ alternative specifications of Equation~\ref{full_spec}. The main specification is based on Equation~\ref{full_spec} which is at the {\SpecChartMainEstimatePercentile} percentile ranking of our income elasticity coefficient estimates from smallest to largest. The 95 percent confidence interval estimates are included and results are plotted from smallest ({\SpecChartMinimumEstimate}) to largest ({\SpecChartMaximumEstimate}) coefficient estimate on $ln(INCOME)$. }
   	\end{minipage}
   	\vspace{0.32cm}
\end{figure}

\FloatBarrier

  \FloatBarrier
 \begin{figure}[t]
   	\vspace{0.32cm}
	\caption{Income elasticity of WTP estimates based on alternative statistical models.}
	\label{fig:appendix_elasticity_models}
	\centering
	\begin{minipage}{1\textwidth} 
	    \includegraphics[width=1\linewidth]{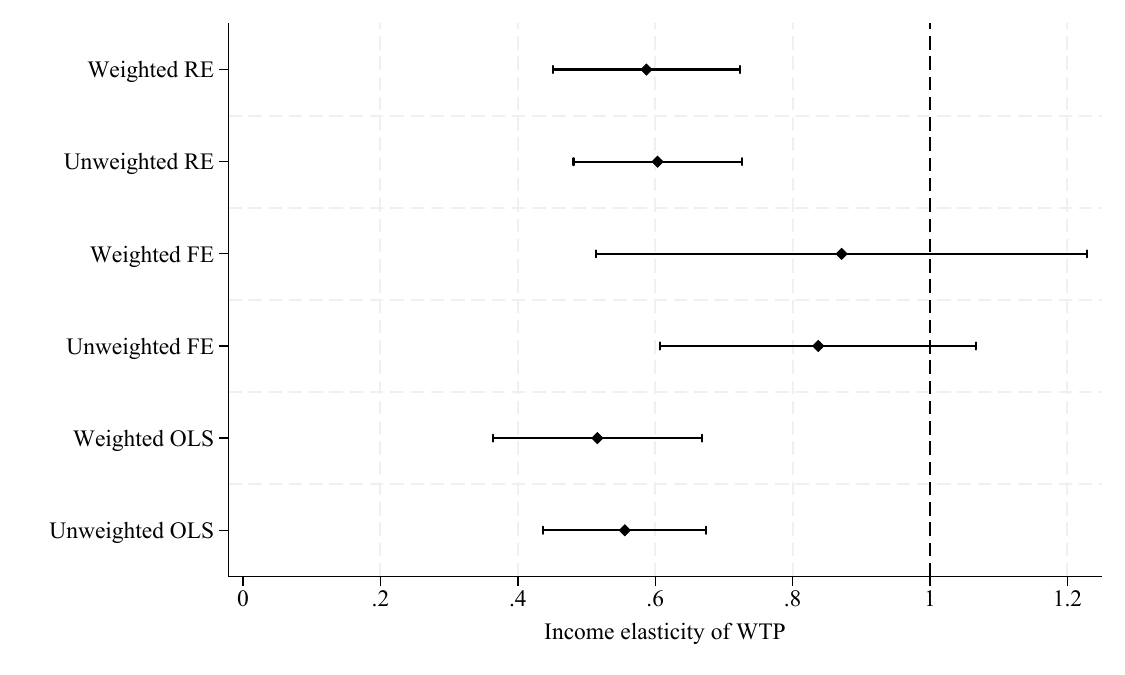 }
	\end{minipage}
 	\begin{minipage}{0.97\textwidth} 
   \flushleft{\footnotesize \emph{Notes:} The main result is based on a random-effects (RE) model weighted by the square root of the sample size. Some frequent alternatives to this approach include unweighted random effects, and both OLS and fixed effects models that are weighted and unweighted. While a Hausman test suggests RE model is most appropriate, we provide these alternative estimates.}
   	\end{minipage}
   	\vspace{0.32cm}
\end{figure}

We find that the regression model chosen also impacts results. Our main specification utilises a random-effects model which also falls between the fixed effects and OLS (and between effects) estimators. However, the fixed-effects alternative would also be derived with substantially less data as it excludes and singleton estimate studies.

\FloatBarrier
\FloatBarrier
We also find an impact from choosing alternative weighting schemes. Follow common convention, we prefer the square-root of the sample size of studies to develop as weights. This places more weight on larger studies but without allowing large studies to entirely dominate the results. In effect, this results in a smaller estimate of the income elasticity of WTP and subsequently a more conservative RPC estimate.

 \begin{figure}[t]
   	\vspace{0.32cm}
	\caption{Income elasticity of WTP estimates by weight selection.}
	\label{fig:appendix_elasticity_weights}
	\centering
	\begin{minipage}{1\textwidth} 
	    \includegraphics[width=1\linewidth]{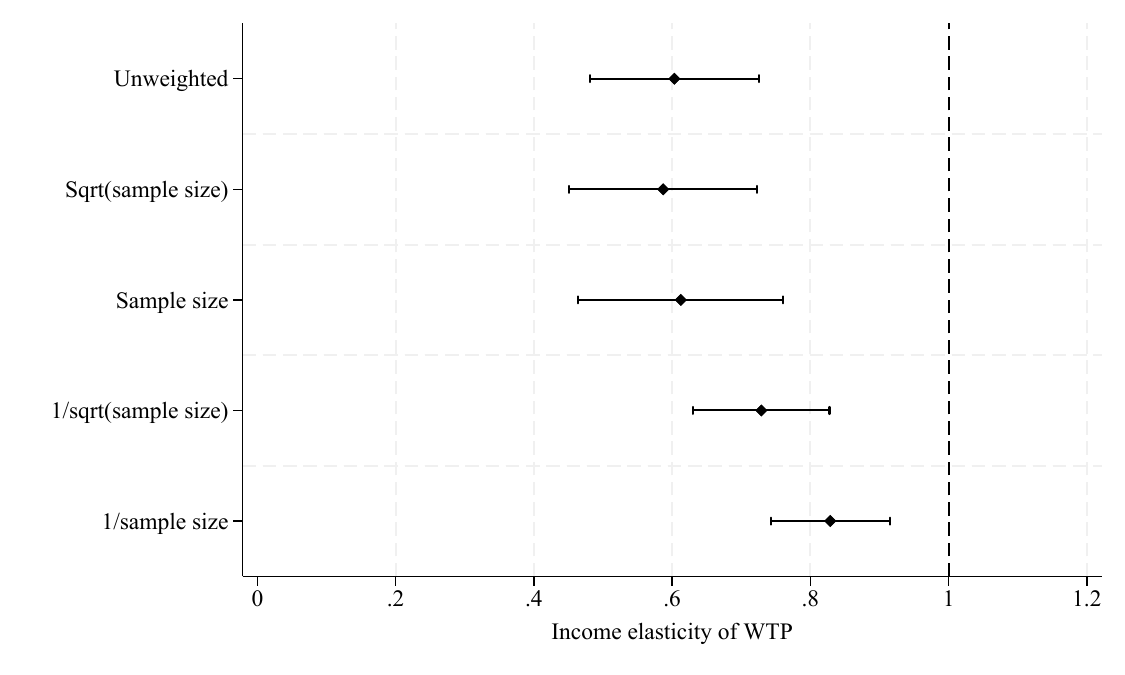 }
	\end{minipage}
 	\begin{minipage}{0.97\textwidth} 
   \flushleft{\footnotesize \emph{Notes:} The main result is derived with weights based on the square root of the sample size. Some alternatives that are more or less reasonable are to use the sample size, inverse of the sample size, and inverse of the square root of the sample size. Inverse sample sizes will tend to place more weight on studies with smaller sample sizes and squared sample size weights will tend to bias estimates toward studies with substantially larger samples. }
   	\end{minipage}
   	\vspace{0.32cm}
\end{figure}

\FloatBarrier

We drop successively larger portions of the dataset at the top and bottom ends to test the sensitivity to outliers. We take this exercise through all top (bottom) income levels from the top (bottom) 1-percent through the bottom (top) 90-percent. We find that in both cases, a larger share - about one-third - of the data would have to be dropped before markedly different income elasticity of willingness to pay estimates would be arrived at.

\begin{figure}[h!]
   	\vspace{0.32cm}
	\caption{Income elasticity of WTP estimates when dropping top income observations.}
	\label{fig:drop-top}
	\centering
	\begin{minipage}{1\textwidth} 
	    \includegraphics[width=1\linewidth]{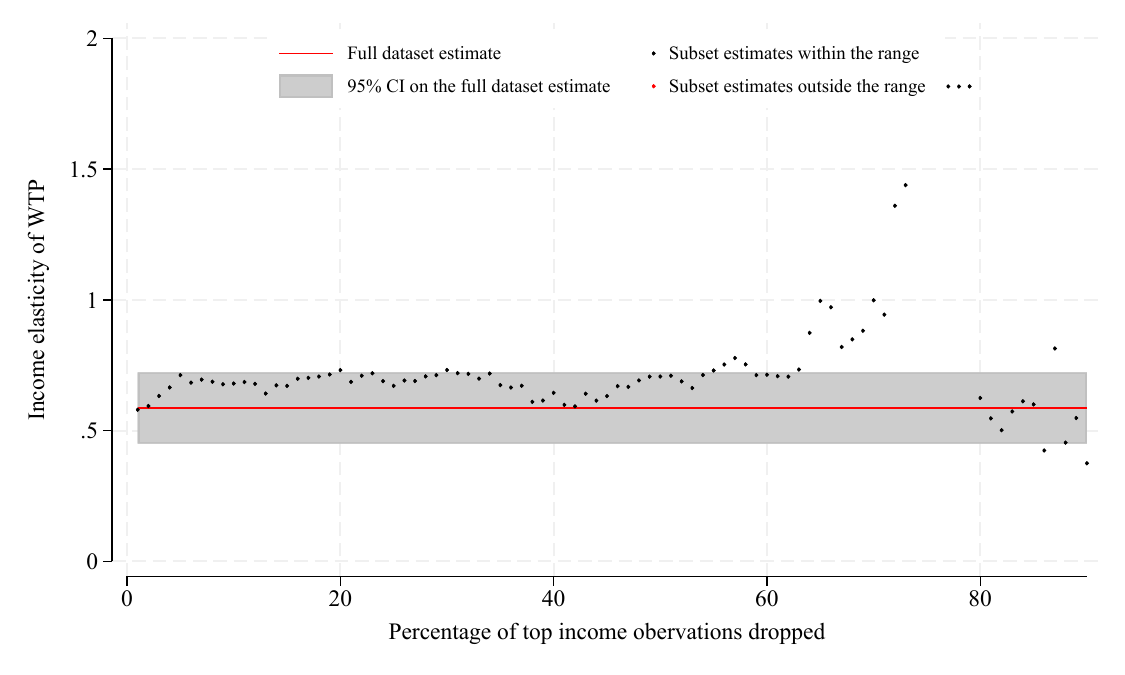 }
	\end{minipage}
 	\begin{minipage}{0.97\textwidth} 
   \flushleft{\footnotesize \emph{Notes:} Alternative estimates are the result of dropping a successively larger share of observations ranked by respondent income. We first drop observations with the top 1-percent of incomes, then proceed in 1-percent increments, dropping the top 2-percent, then 3-percent, and so on up to 90-percent.}
   	\end{minipage}
   	\vspace{0.32cm}
\end{figure}

\FloatBarrier

\begin{figure}[h!]
   	\vspace{0.32cm}
	\caption{Income elasticity of WTP estimates when dropping bottom income observations.}
	\label{fig:drop-bottom}
	\centering
	\begin{minipage}{1\textwidth} 
	    \includegraphics[width=1\linewidth]{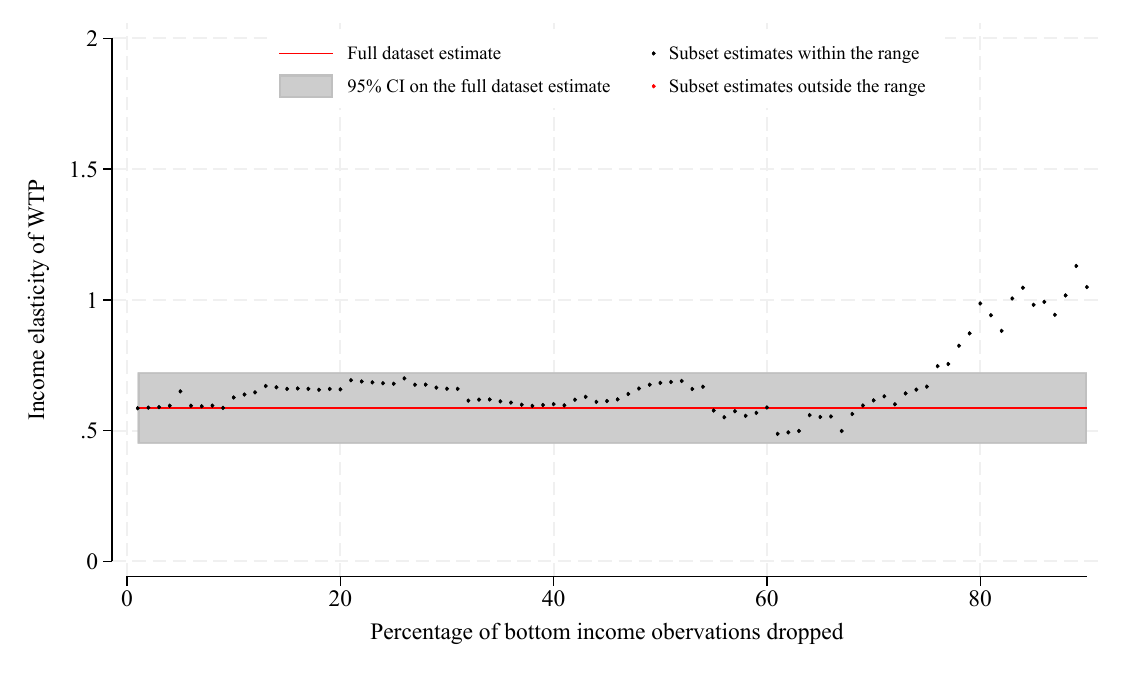 }
	\end{minipage}
 	\begin{minipage}{0.97\textwidth} 
   \flushleft{\footnotesize \emph{Notes:} Alternative estimates are the result of dropping a successively larger share of observations ranked by respondent income. We first drop observations with the bottom 1-percent of incomes, then proceed in 1-percent increments, dropping the bottom 2-percent, then 3-percent, and so on.}
   	\end{minipage}
   	\vspace{0.32cm}
\end{figure}

\FloatBarrier

Finally, we randomly drop successively larger shares of our dataset to test whether subsets are driving our results. We randomly select and drop data 25 times at each percentage level, from dropping 1-percent of the data through 90-percent - 2,250 draws in total. As in top- and bottom-income dropping exercises, we find that a substantial share of the data must be dropped to substantially impact the results.

\begin{figure}[h!]
   	\vspace{0.32cm}
	\caption{Sensitivity of income elasticity of WTP estimates to randomly dropping data.}
	\label{fig:drop-random}
	\centering
	\begin{minipage}{1\textwidth} 
	    \includegraphics[width=1\linewidth]{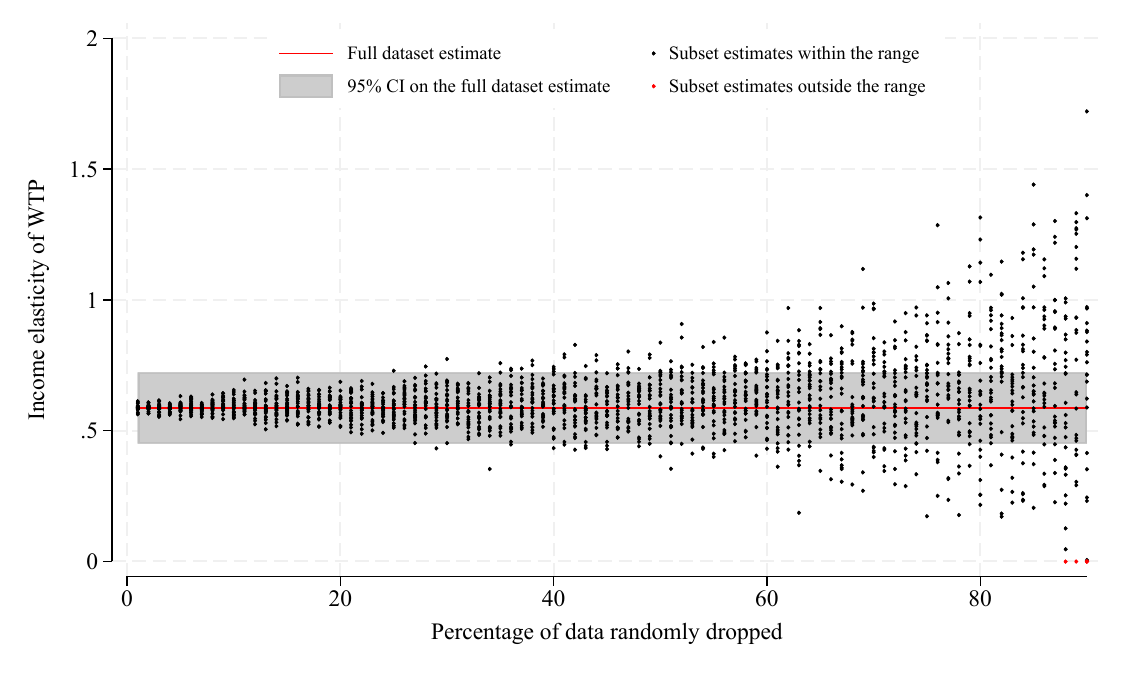 }
	\end{minipage}
 	\begin{minipage}{0.97\textwidth} 
   \flushleft{\footnotesize \emph{Notes:} Alternative estimates based on dropping random draws of the dataset. At each 1-percent increment, 25 draws of the data are used to re-estimate the income elasticity of WTP using our main model specification. Data is randomly drop in successively larger amountds from 1-percent through 90-percent of the data in 1-percent increments.}
   	\end{minipage}
   	\vspace{0.32cm}
\end{figure}

\FloatBarrier

\end{appendices}

\end{document}